\documentclass[apj]{emulateapj}
\usepackage{graphicx}
\usepackage{psfig}

\shorttitle{Active Region Plasma Diagnostics with Hinode}
\shortauthors{Testa et al.}
\usepackage{color} 
\usepackage{amsmath}           
\usepackage{amsfonts}

\def \arcsec {\hbox{$^{\prime\prime}$}}

\def \ll   {$\lambda$}
\def \ergs {erg~s$^{-1}$}
\def \flxu {erg~cm$^{-2}$~s$^{-1}$}

\def \emt {$EM(T)$}

\def \hinode   {{\em Hinode}}
\def \xrt {{\sc XRT}}
\def \eis  {{\sc EIS}}

\def \feviii  {Fe\,{\sc viii}}
\def \feix    {Fe\,{\sc xix}}
\def \fex     {Fe\,{\sc x}}
\def \fexi    {Fe\,{\sc xi}}
\def \fexii   {Fe\,{\sc xii}}
\def \fexiii  {Fe\,{\sc xiii}}
\def \fexiv   {Fe\,{\sc xiv}}
\def \fexv    {Fe\,{\sc xv}}
\def \fexvi   {Fe\,{\sc xvi}}
\def \fexvii  {Fe\,{\sc xvii}}

\def \fexxii  {Fe\,{\sc xxii}}
\def \fexxiii {Fe\,{\sc xxiii}}
\def \fexxiv  {Fe\,{\sc xxiv}}
\def \nixvii  {Ni\,{\sc xvii}}
\def \sx     {S\,{\sc x}}
\def \sxiii   {S\,{\sc xiii}}
\def \sivi    {Si\,{\sc vi}}
\def \sivii   {Si\,{\sc vii}}
\def \six     {Si\,{\sc x}}
\def \caxvi   {Ca\,{\sc xvi}}
\def \heii   {He\,{\sc ii}}
\def \ov     {O\,{\sc v}}

\def\ion[#1 #2]{#1\,{\sc #2}}
\def\ergs[#1]{#1 {ergs}~{cm$^{-2}$}\,{s$^{-1}$}\,{sr$^{-1}$}}
\def\dens[#1]{10$^{#1}$\hskip 1.5pt{cm$^{-3}$}}
\def\densr[#1 #2]{10$^{#1}$\hskip 1pt{--}\hskip .5pt{10$^{#2}$}\hskip 1.5pt{cm$^{-3}$}}
\def\fl[#1 #2]{{#1}$\pm${#2}}
\def\orb[#1 #2]{{$#1^{#2}$}}
\def\ls[#1 #2]{{$^{#1}${#2}}}
\def\tm[#1 #2 #3]{{$^{#1}${#2}$_{#3}$}}

\newcounter{ion}

\begin{document}
\title{Temperature distribution of a non-flaring active region from simultaneous 
	\hinode\ \xrt\ and \eis\ observations}
\author{Paola Testa\altaffilmark{1}, Fabio Reale\altaffilmark{2,3}, 
Enrico Landi\altaffilmark{4}, Edward E.\ DeLuca\altaffilmark{1}, Vinay
Kashyap\altaffilmark{1}}
\altaffiltext{1}{Smithsonian Astrophysical Observatory,
	60 Garden street, MS 58, Cambridge, MA 02138, USA; 
	ptesta@cfa.harvard.edu}
\altaffiltext{2}{Dipartimento di Scienze Fisiche ed Astronomiche,
  Sezione di Astronomia, Universit\`a di Palermo, Piazza del
  Parlamento 1, 90134, Italy}
\altaffiltext{3}{INAF-Osservatorio Astronomico di Palermo, Piazza del
  Parlamento 1, 90134 Palermo, Italy}
\altaffiltext{4}{Department of Atmospheric, Oceanic and Space
  Sciences, University of Michigan 2455 Hayward St., Ann Arbor MI
  48109 USA}

\begin{abstract}

We analyze coordinated \hinode\ \xrt\ and \eis\ observations of a 
non-flaring active region to investigate the thermal properties of 
coronal plasma taking advantage of the complementary diagnostics 
provided by the two instruments. 
In particular we want to explore the presence of hot plasma 
in non-flaring regions.  
Independent temperature analyses from the \xrt\ multi-filter dataset,
and the \eis\ spectra, including the instrument entire wavelength 
range, provide a cross-check of the different temperature 
diagnostics techniques applicable to broad-band and spectral data 
respectively, and insights into cross-calibration of the two 
instruments.

The emission measure distribution, \emt, we derive from the two 
datasets have similar width and peak temperature, but show a 
systematic shift of the absolute values, the \eis\ \emt\ being 
smaller than \xrt\ \emt\ by approximately a factor 2.  We explore 
possible causes of this discrepancy, and we discuss the influence 
of the assumptions for the plasma element abundances. Specifically,
we find that the disagreement between the results from the two
instruments is significantly mitigated by assuming chemical
composition closer to the solar photospheric composition rather than
the often adopted ``coronal'' composition \citep{Feldman92}.

We find that the data do not provide conclusive evidence on the 
high temperature ($\log T[{\rm K}] \gtrsim 6.5$) tail of the plasma 
temperature distribution, however, suggesting its presence to a level
in agreement with recent findings for other non-flaring regions.

\end{abstract}


\section{Introduction}
\label{s:intro}

Understanding how solar and stellar coronae are heated to high temperatures 
is one of the most important open issues in astrophysics. Coronal heating
is clearly related to the strong magnetic fields that fill the atmospheres
of the Sun and solar-like stars, but the mechanism that converts magnetic 
energy into thermal energy remains unknown. Theoretical models based on 
steady heating failed to explain the physical properties observed in coronal
structures. A promising and widely studied framework for understanding 
coronal heating is the nanoflare model proposed by \citet{Parker72}. In this 
model, convective motions in the photosphere lead to the twisting and 
braiding of coronal magnetic field lines. This topological complexity 
ultimately leads to the formation of current sheets, where the magnetic 
field can be rearranged through the process of magnetic reconnection. 
This model has been further refined and adapted to a scenario where nanoflares
occur in unresolved strands in coronal loops, below the spatial resolution
of available instrumentation 
\citep{Parker88,Cargill94,Cargill97,Klimchuk06,Warren03,Parenti06}.

Nanoflare models predict the presence of very hot plasma with temperatures
in excess of 3~MK in non flaring solar regions; depending on the energy
of the nanoflare, temperatures may reach or even exceed 10~MK. Unambiguous
detection of such extreme temperatures in non-flaring solar regions can
provide convincing evidence for the presence of nanoflares. However, such
detection is not easy, as the amount of very hot plasma produced by
nanoflares is expected to be very small. Recent studies have provided 
some evidence of the presence of hot plasma in the non-flaring Sun.
\citet{McTiernan09} carried out an analysis of the temperature and emission 
measure determined from quiescent plasma during the 2002-2006 decay phase 
of solar cycle 23, as derived from GOES and RHESSI observations. 
He found a persistent faint plasma component with temperatures approximately
constant during the entire 2002-2006 interval, and approximately between 5 
and 10~MK. However, GOES and RHESSI provided different values of such 
temperature, and their results were not necessarily well correlated; also, 
this analysis relied on the isothermal plasma assumption.

Other studies tried to identify the hot plasma through the X-ray emission 
observed by the X-ray Telescope (\xrt; \citealp{Golub07}) onboard \hinode\ 
\citep{Kosugi07}. 
\citeauthor{Reale09} carried out a temperature analysis of a 
non-flaring active region, first with only \xrt\ multi-filter data 
\citep{Reale09} and then combining \xrt\ and RHESSI observations 
\citep{Reale09b}. Their findings point to the presence of small 
amounts of very hot plasmas, with temperatures of $\simeq 5-10$~MK,
and emission measure of the order of few percent of the dominant
cool component. These characteristics of the emission measure
distribution are compatible with the predictions of nanoflare models. 
However, they spelled out and discussed the main limitations of 
their study and similar analyses. First, \xrt\ is also sensitive 
to plasma at normal active region temperatures (2-3~MK) and thus 
contamination from the colder active region plasma is a considerable 
obstacle to the detection of the much smaller amounts of hot plasma; 
also, the limited temperature resolution of \xrt\ prevents a detailed
study of the cold component. Second, RHESSI sensitivity makes 
it very hard to even detect the quiescent active region plasma. 
Third, instrumental calibration is an issue for both instruments.
\citet{Schmelz09a} also detected a faint hot temperature tail to the 
emission measure distribution of active region plasma, and 
determined its temperature to be around 30~MK. Subsequent analyses 
by \citet{Schmelz09b} included RHESSI data and, while confirming the 
presence of such hot material, could not reconcile the \xrt\ and 
RHESSI observations using the standard calibration of both instruments.
A self-consistent solution was only found if a series of instrumental 
parameters and the plasma element abundances were adjusted, and the 
temperature of the hot plasma decreased. 

Ample efforts have been devoted to the accurate determination of the 
thermal structuring of coronal plasma to derive robust
observational constraints to the mechanism(s) of coronal heating.
The plasma temperature distribution of the quiet corona and of 
active regions has been investigated through imaging data
and spectroscopic observations \citep[e.g.,][]{Brosius96,Landi98,
Aschwanden00,Testa02,DZM03,Reale07,Landi09,Shestov10,Sylwester10}.
Several recent works have focused on EUV spectra obtained with the 
\hinode\ Extreme Ultraviolet Imaging Spectrometer (\eis; \citealt{Culhane07} )
which provides good temperature diagnostic capability, together with
higher spatial resolution and temporal cadence than previously 
available \citep[e.g.,][]{Watanabe07,Warren08loops,Patsourakos09,
Brooks09,Warren09}.

In the present work, we address the issue of determining the temperature
distribution of coronal plasma from a different perspective:
we investigate thermal properties of coronal plasma in non-flaring 
active regions using simultaneous \hinode\ observations with \xrt\ and
with \eis,  which provide complementary diagnostics for the X-ray emitting 
plasma. The multi-filter \xrt\ dataset together with \eis\ spectra, 
including its entire wavelength range, allow to accurately 
determine the thermal structure of the active region plasma, and 
to explore the presence of hot plasma in non-flaring regions.  
We use spectroscopic observations from the \hinode/\eis\ instrument 
of a quiescent active region to constrain the emission measure distribution
of the bulk of the active region plasma with the spectral lines observed by 
\eis\ in the 171-212\AA\ and 245-291\AA\ spectral ranges (see also
e.g., \citealt{Young07,Doschek07}). 
Since \eis\ is most sensitive to plasma with temperatures of 
0.6-2~MK, \eis\ spectra allow us to accurately determine the emission
measure distribution of the quiescent active region plasma, to 
evaluate the fraction of the observed \xrt\ count rates that it emits, 
and thus investigate the true amount of emission from the nanoflaring 
plasma. Thus, the combination of \xrt\ and \eis\ observations of the 
same active region allows us to characterize the plasma temperature 
distribution with better detail than in previous studies. 
In fact, while some previous studies have made use 
of data from both imaging and spectroscopic data to constrain the 
properties of  the emitting plasma \citep[e.g.,][]{Warren10,Landi10,ODwyer10},
to our knowledge no previous work has carried out a determination 
of the temperature distribution by combining \xrt\ and \eis\ data, 
nor a quantitative comparison of the independent analysis from the 
different instruments, as we do here.
Independent temperature analysis from the two datasets provide 
a cross-check of the different temperature diagnostics techniques 
applicable to spectral and broad-band data respectively, and 
insights into cross-calibration of the two instruments.

The observations are described in Section~\ref{s:obs}. The data analysis and
results of the determination of the plasma temperature distribution are
presented in Section~\ref{ss:results}. Our findings are discussed in 
Section~\ref{s:discuss} and summarized in Section~\ref{s:conclusions}.

\section{Observations}
\label{s:obs}

\begin{deluxetable*}{cccccccc} 
 \small
\tablecolumns{8} 
\tablewidth{0pc} 
\tablecaption{Details of \hinode\ \xrt\ and \eis\ observations analyzed 
	in this paper, and shown in Figures~\ref{fig:obs_xrt}, 
	and \ref{fig:obs_eis}.
	\label{tab:obs}} 
\tablehead{ 
\colhead{}  & \multicolumn{6}{c}{\xrt}   & \colhead{\eis} \\ 
\cline{2-7} 
\colhead{}  & \colhead{Al-poly}  & \colhead{C-poly}  & \colhead{Ti-poly}  & 
  \colhead{Be-thin}  & \colhead{Be-med}  & \colhead{Al-med}  & \colhead{171-212\AA, 245-291\AA} }
\startdata 
 
FOV               & \multicolumn{6}{c}{384\arcsec$\times$384\arcsec}  &  128\arcsec$\times$128\arcsec\ \\
START OBS         & \multicolumn{6}{c}{2008-06-20T23:27:30}          & 2008-06-20T23:03:39  \\
END OBS           & \multicolumn{6}{c}{2008-06-21T03:38:50}          & 2008-06-21T02:19:14  \\
$t_{\rm exp}$ [s]  &  4.1 &  8.2 &  8.2 &  23  &  33  &  46         &  90 \\

\enddata 
\end{deluxetable*}

We observed the non-flaring NOAA active region 10999 close to disk center, 
beginning on June 20 2008 around 23~UT for several hours with both the 
X-ray Telescope and the Extreme Ultraviolet Imaging Spectrometer onboard 
\hinode. 
The details of the observations are presented in Table~\ref{tab:obs}.  

\begin{figure*}[!ht]
\centerline{
\includegraphics[scale=1.2]{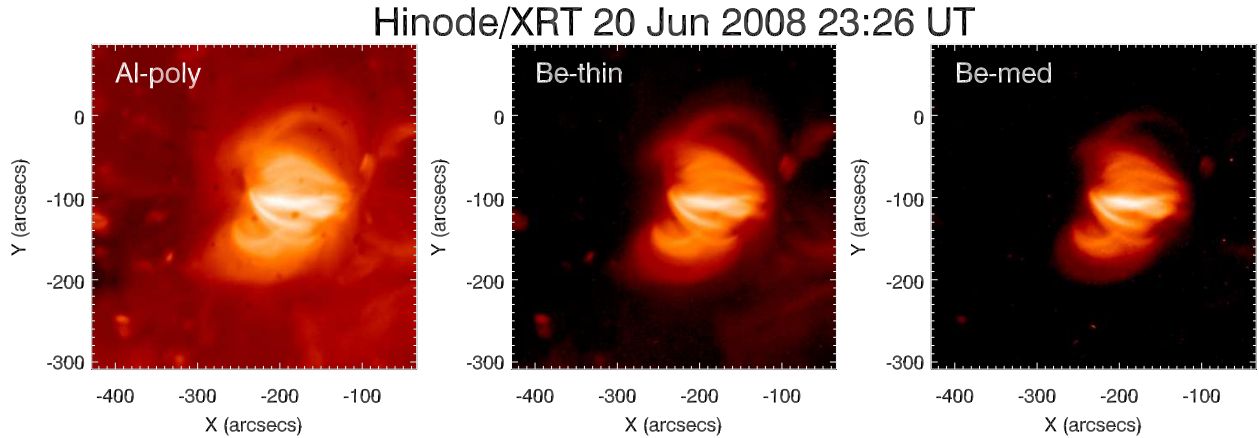}
}
\caption{ \xrt\ images of AR~10999, obtained summing all the images 
	taken in a given filter over the $\sim 3$~hr observation
        (see Table~\ref{tab:obs}): Al-poly  
	({\em left}; total integration time $t_{\rm int} \sim 131$~s), 
	Be-thin ({\em center}; $t_{\rm int} \sim 828$~s), and
	Be-med filter ({\em right}; $t_{\rm int} \sim 1155$~s). 
	\label{fig:obs_xrt}}
\end{figure*}

\xrt\ observed AR 10999 in several filters, with a field of view (FOV) 
of 384\arcsec$\times$384\arcsec, for about 4 hours starting June 20 2008 
at 23:27~UT, with a cadence of about 5~minutes in each filter. 
We analyze observations in the following filters: Al\_poly, C\_poly, 
Ti\_poly, Be-thin, Be-med, Al-med. The \xrt\ data were processed with 
the standard routine xrt\_prep, available in SolarSoft to remove the 
CCD dark current, and cosmic-ray hits.
Figure~\ref{fig:obs_xrt} shows the images of \xrt\ observations in 
Al-poly, Be-thin and Be-med, integrated over the entire observing 
time (see also Table~\ref{tab:obs} for details).

\begin{figure*}[!ht]
\centerline{
\includegraphics[angle=90,scale=0.7]{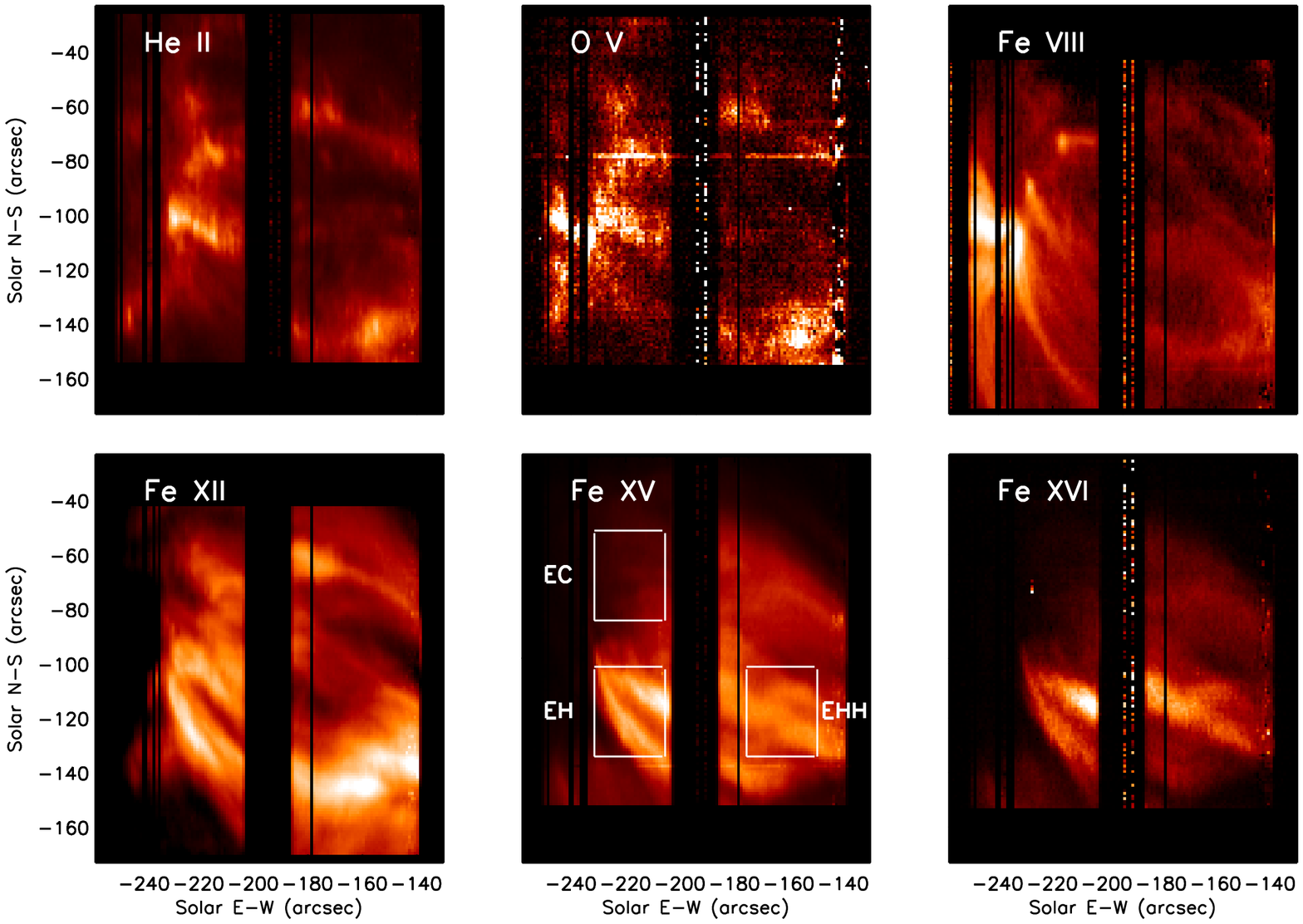}
\vspace{-0.5cm}}
\caption{\eis\ images in spectral lines formed at different
	characteristic plasma temperatures: \heii\ (256.32\AA, 
	$\log(T_{\rm max}[{\rm K}])=4.9$), \ov\ (248.46\AA, 
	$\log(T_{\rm max}[{\rm K}])=5.4$), \feviii\ (185.21\AA, 
	$\log(T_{\rm max}[{\rm K}])=5.55$), \fexii\ (195.12\AA, 
	$\log(T_{\rm max}[{\rm K}])=6.15$), \fexv\ (284.16\AA, 
	$\log(T_{\rm max}[{\rm K}])=6.35$), \fexvi\ (262.98\AA, 
	$\log(T_{\rm max}[{\rm K}])=6.4$). 
	In the 284\AA\ \fexv\ image we indicate the three regions selected
	for the analysis of the plasma thermal structure (see 
	Figure~\ref{fig:tmap} and \S\ref{ss:results}): EC ({\em top left}),
	EH ({\em bottom left}), EHH ({\em bottom right}).
	\label{fig:obs_eis}} 
\end{figure*}

The FOV of the \eis\ observations analyzed here is 
128\arcsec$\times$128\arcsec\ and was built up by stepping the 
1\arcsec\ slit from solar west to east over a 3.5 hr period 
from 23:03~UT of June 20th to 02:19~UT of June 21st. 
The study includes full spectra on both the \eis\ detectors from 
171-212\AA\ and 245-291\AA. The exposure time was 90~s and the 
study acronym is HPW001\_FULLCCD\_RAST. 
The observations were carried out during eclipse season and they 
were not paused during eclipses: the black stripes of missing data 
in Figure~\ref{fig:obs_eis} correspond to \hinode\ eclipses.
The \eis\ data are processed with the eis\_prep routine available in 
SolarSoft to remove the CCD dark current, cosmic-ray strikes on the 
CCD, and take into account hot, warm, and dusty pixels. In addition, 
the radiometric calibration is applied to convert the data from 
photon events to physical units. 
The \eis\ routine eis\_ccd\_offset is then used to correct for the 
wavelength dependent relative offset of the two CCDs of 1-2~pixels 
in the X-direction, and $\sim 18$~pixels in the Y-direction. 
Figure~\ref{fig:obs_eis} shows images obtained from \eis\ observations 
by integrating over narrow wavelength ranges each dominated by a 
single line with different characteristic temperature of formation. 
We also show three small areas of the active region which have 
been selected for the detailed analysis of thermal structuring 
(see next section for details).

\section{Analysis Methods and Results}
\label{ss:results}

\begin{figure}[!ht]
\centerline{\includegraphics[scale=0.5]{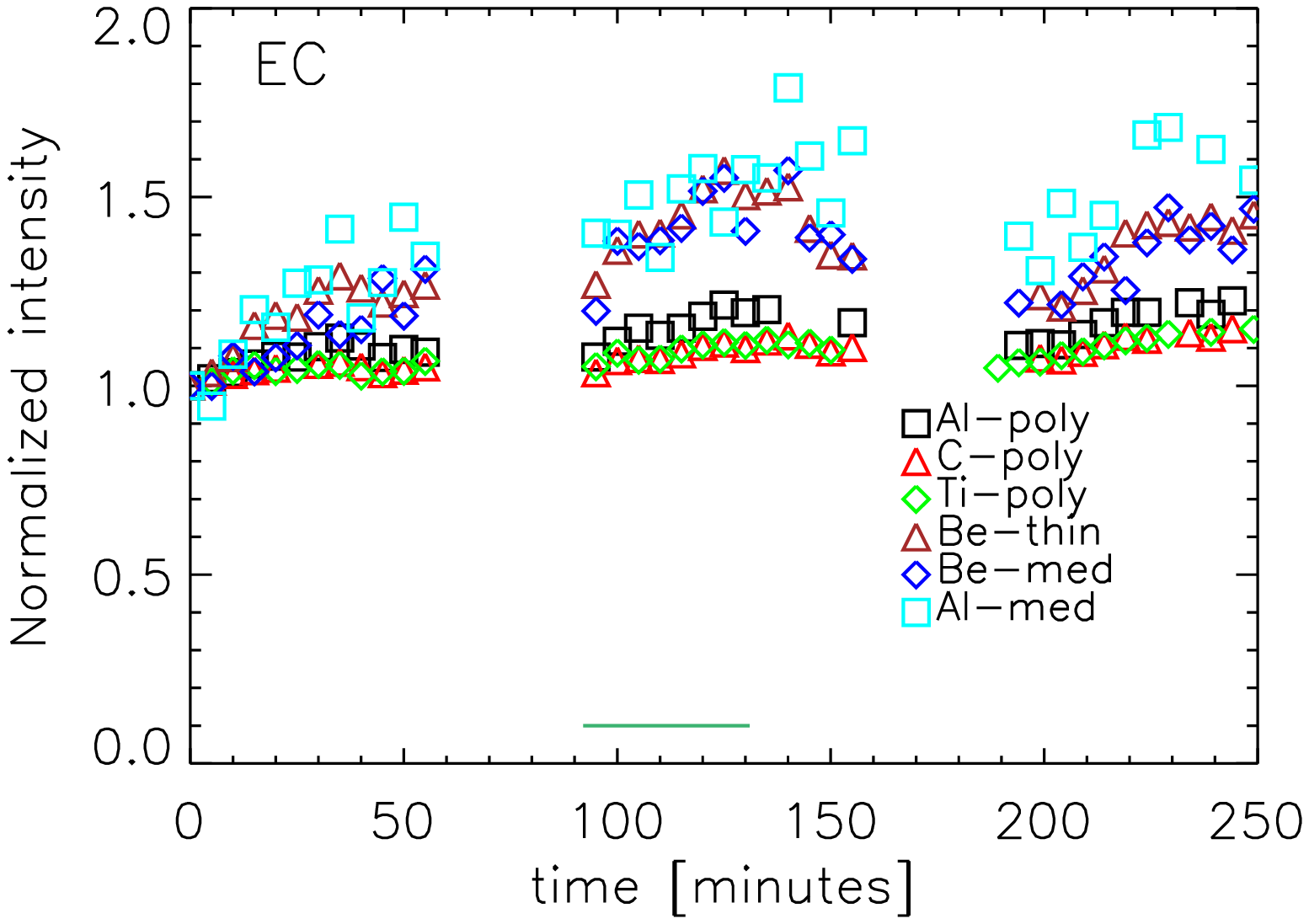}\vspace{-0.3cm}}
\centerline{\includegraphics[scale=0.5]{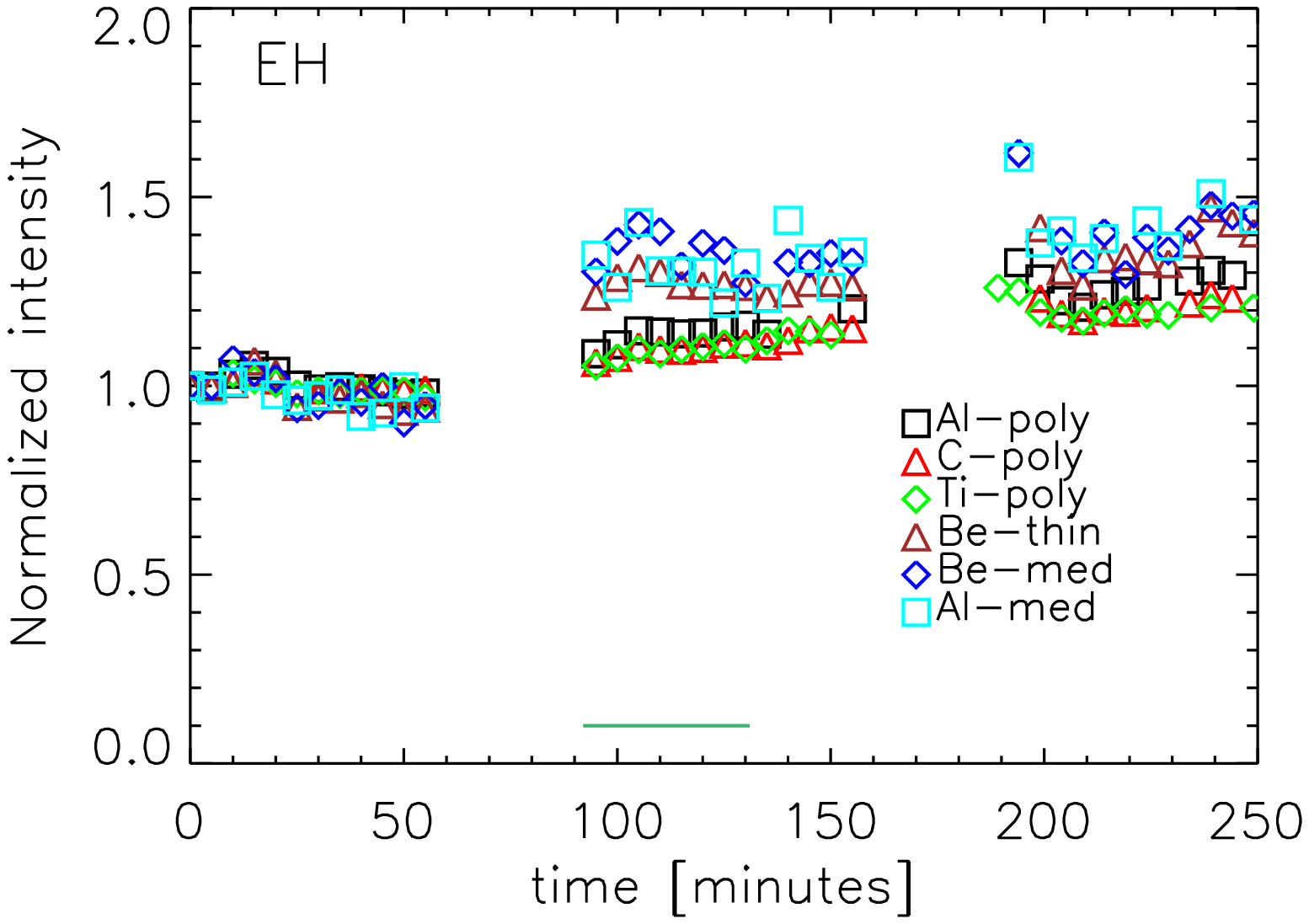}\vspace{-0.3cm}}
\centerline{\includegraphics[scale=0.5]{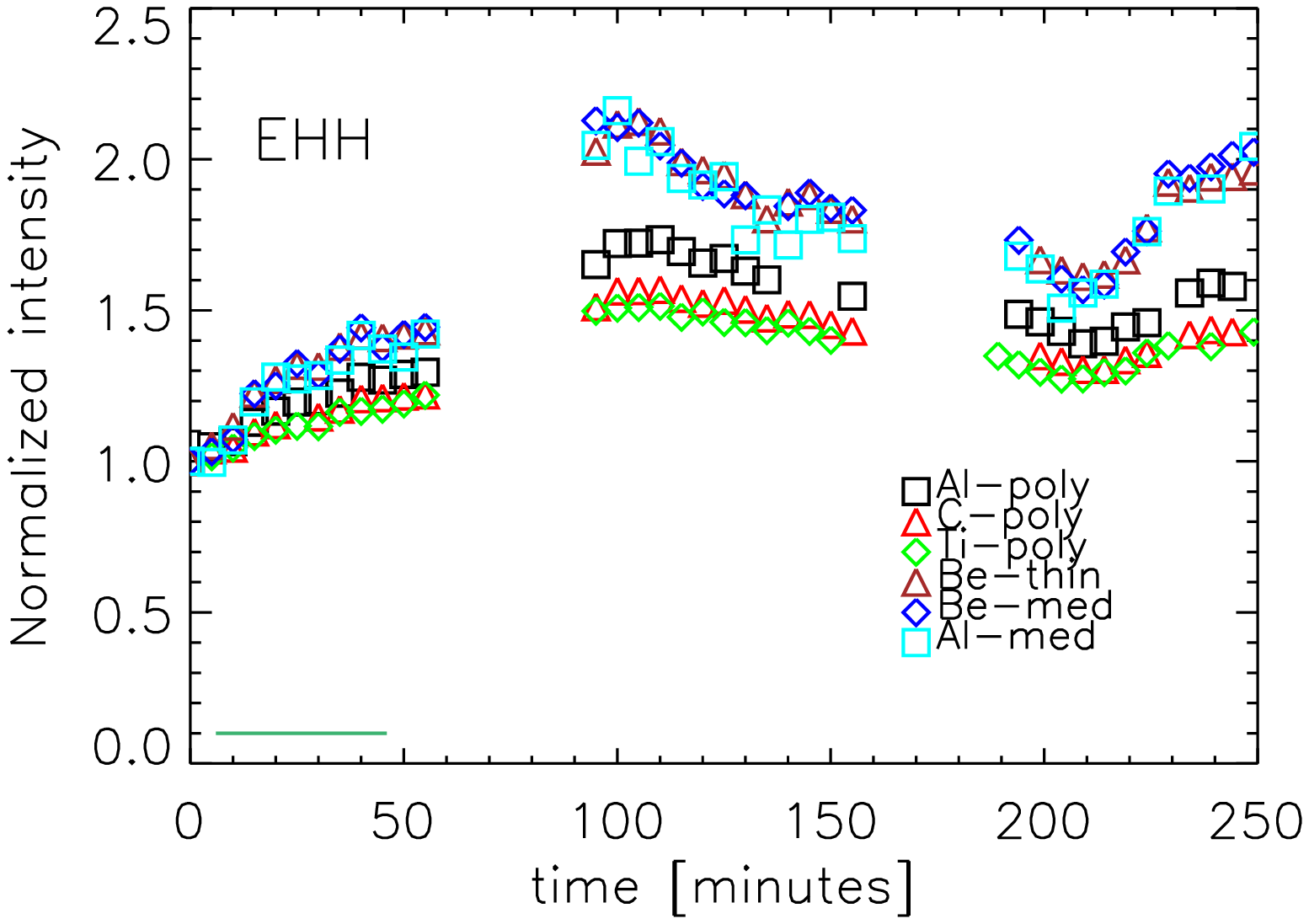}\vspace{-0.3cm}}
\caption{Lightcurves of \xrt\ observations in all analyzed filters, for the three
	selected regions:  EC ({\em top}), EH ({\em middle}), EHH ({\em bottom}).
	For each filter the signal is normalized to the intensity in
        the first image. 
        The solid line at the bottom of the plot indicates the time
        interval when EIS was observing the selected region.
	\label{fig:lc_xrt}}
\end{figure}

Inspection of the \xrt\ observations, in all filters, indicate that 
the active region is characterized by a modest level of variability 
over a wide range of temperatures (see Figure~\ref{fig:lc_xrt}). 
Therefore, in order to increase S/N, we have analyzed the \xrt\ 
dataset obtained by coaligning the images taken in each filter at 
different times, and then summing them up. 

\begin{figure}[!ht]
\vspace{-4.5cm}
\centerline{
\includegraphics[scale=0.6]{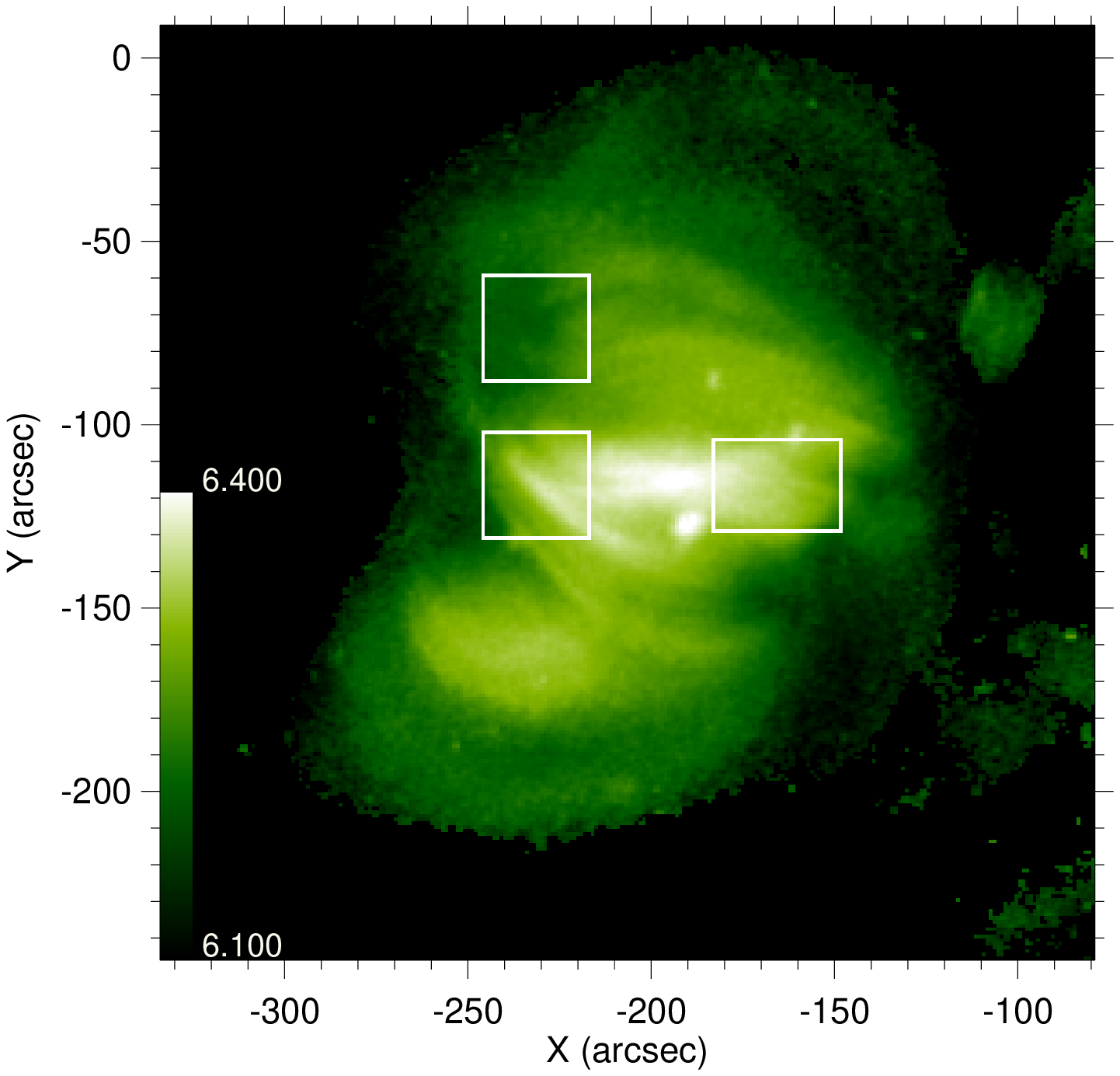}
}
\vspace{-0.8cm}
\caption{ Temperature map obtained from the combined improved filter 
	ratio from the \xrt\ images of AR~10999 shown in Figure~\ref{fig:obs_xrt}.
	The three regions selected for the detailed temperature analysis
	are indicated: EC ({\em top left}), EH ({\em bottom left}), EHH 
	({\em bottom right}).
	When selecting the boundaries of region EHH we modified the shape
	(however maintaining the area value) in order to avoid the 
	contamination spots clearly visible in the temperature map 
	as bright blobs immediately to the east and north of the 
	selected area.
	\label{fig:tmap}}
\end{figure}

In order to carry out a direct and detailed comparison of thermal 
analysis from \xrt\ and \eis\ data, we have then selected a few 
subregions. 
To select these regions we have first obtained maps of temperature 
and emission measure over the whole active region, by using estimates 
derived with the so-called combined improved filter ratio method, 
devised by \cite{Reale07}. The resulting temperature map is shown in
Figure~\ref{fig:tmap}. We selected three regions, of area approximately 
30\arcsec$\times$30\arcsec: EC, which in the temperature map
homogeneously appears as relatively cool, and two hotter regions
(EH, EHH).  The values of temperature mean and standard
deviation derived for these three regions from the CIFR diagnostics
are: $1.5 \pm 0.1$~MK (EC), $1.9 \pm 0.2$~MK (EH), and $2.0 \pm
0.2$~MK (EHH). 
When selecting the boundaries of region EHH in the \xrt\ 
data, we slightly modified the shape (however maintaining the area 
value) in order to avoid the contamination spots clearly visible in 
the temperature map as bright blobs immediately to the east and north
of the selected area.
Figure~\ref{fig:obs_eis} shows the selected regions in the \eis\ 
\fexv\ 284\AA\ image. 
For the temperature analysis presented in the rest of the paper we 
integrated the \xrt\ and \eis\ signals over each of these regions,
over the time intervals when the two instruments were simultaneously
observing the selected region. Figure~\ref{fig:lc_xrt} shows that the
\xrt\ light curves of each region changed little over the time period
when \eis\ was observing the same areas.

\begin{figure}[!ht]
\centerline{
\includegraphics[scale=0.5]{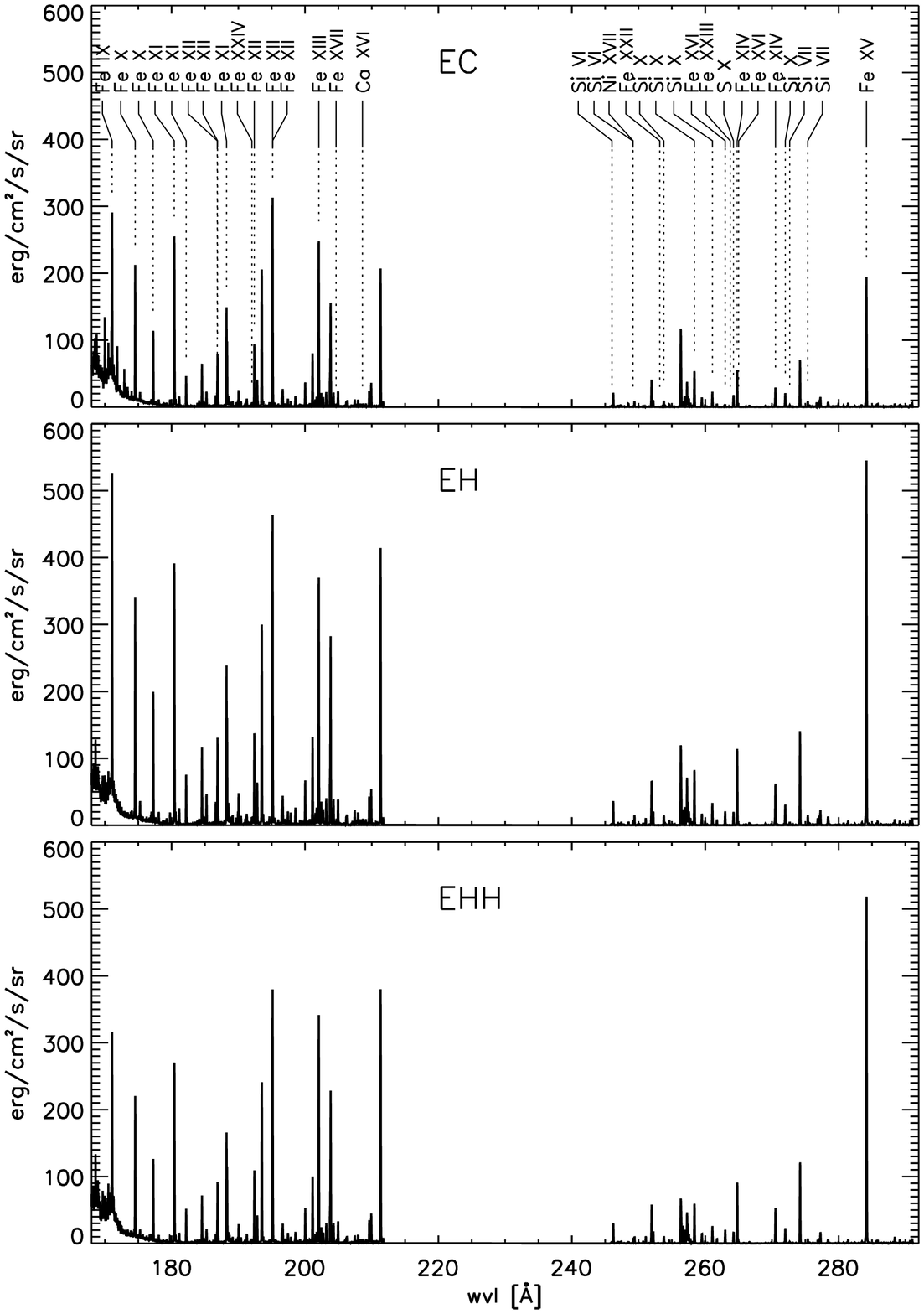}
\vspace{-3cm}}
\caption{\eis\ spectra of the three regions selected for the analysis:
	EC ({\em top}), EH ({\em middle}), EHH ({\em bottom}).
	The identification of the spectral lines used for the analysis
	of the plasma thermal properties (listed in 
	Table~\ref{tab:eislines}) are shown in the spectrum in the 
	top panel.
	\label{fig:spec_eis}}
\end{figure}

Figure~\ref{fig:spec_eis} shows the \eis\ full spectra in the three 
regions selected for the analysis, with the identification of the 
brightest lines. The \eis\ exposure time (90~s at each location) 
yields high signal in the strongest lines, mostly produced in a 
temperature range $\log T[{\rm K}] \sim 5.8-6.3$. 
Hotter lines present in the \eis\ wavelength range have generally low 
intensities, in typical solar non-flaring conditions, and are 
difficult to detect. We have selected a list of lines which are 
suitable for an accurate derivation of the plasma thermal properties. 
This list (see Table~\ref{tab:eislines}) includes strong lines, 
unblended and with reliable atomic data, and we also included hot 
lines for which however we can only derive upper limits from the 
\eis\ observations of this non-flaring active region. These upper 
limits are nevertheless useful to constrain the high temperature 
component ($\log T[{\rm K}] \gtrsim 6.5$). 

\begin{deluxetable*}{ccccccl} 
\tablecolumns{7} 
\tablewidth{0pc} 
\tablecaption{Identification and measured fluxes of \eis\ spectral 
	lines used for reconstructing the DEM of the three regions 
	selected for the analysis and shown in Figure~\ref{fig:obs_eis}.
	\label{tab:eislines}} 
\tablehead{ 
\colhead{\ll\ [\AA]}  & \colhead{Ion} & 
	\colhead{$\log (T_{\rm max} [{\rm K}])$} &  
	\multicolumn{3}{c}{flux [\flxu]}   & \colhead{Notes\tablenotemark{a}} \\ 
\cline{4-6} 
\colhead{} & \colhead{}   & \colhead{}    & \colhead{EC} & 
\colhead{EH}    & \colhead{EHH}   & \colhead{}}
\startdata 
  
  171.0755   &  \feix     &  5.9   & 8494.52 & 16757.8 & 9238.51 &   \\ 
  174.5340   &  \fex      &  6.0   & 7310.57 & 12152.6 & 7888.11 &   \\ 
  177.2430   &  \fex      &  6.0   & 4231.15 & 7555.12 & 4736.85 &   \\  
  180.4080   &  \fexi     &  6.1   & 10609.4 & 16354.4 & 11523.8 &   \\ 
  182.1690   &  \fexi     &  6.1   & 1784.80 & 3021.49 & 1914.51 &   \\	
  186.8520   &  \fexii    &  6.2   & 694.141 & 1099.68 & 426.001 &   \\    
  186.8840   &  \fexii    &  6.2   & 3129.81 & 5212.59 & 3902.01 &   \\    
  188.2320   &  \fexi     &  6.1   & 5613.69 & 8863.01 & 6469.60 &   \\   
  192.0285   &  \fexxiv   &  7.3   & 310.929 & 539.337 & 304.685 &  u   \\   
  192.3930   &  \fexii    &  6.2   & 4107.59 & 5858.81 & 4550.20 &   \\   
  195.1180   &  \fexii    &  6.2   & 14092.9 & 21301.2 & 16415.0 &   \\   
  202.0440   &  \fexiii   &  6.2   & 11240.2 & 15845.5 & 14868.5 &   \\   
  204.6542   &  \fexvii   &  6.7   & 37.2628 & 60.1331 & 44.5043 &  u   \\   
  208.6040   &  \caxvi    &  6.7   & 5.61904 & 17.4814 & 38.7130 &  u   \\   
  246.0200   &  \sivi     &  5.6   & 76.5847 & 91.3415 & 28.2609 &   \\   
  249.1240   &  \sivi     &  5.6   & 29.2495 & 38.8003 & 15.6522 &   \\   
  249.1780   &  \nixvii   &  6.5   & 20.5272 & 277.752 & 312.949 &   \\    
  253.1702   &  \fexxii   &  7.1   & 3.55755 & 9.38507 & 9.47460 &  u   \\   
  253.7880   &  \six      &  6.1   & 379.199 & 578.696 & 436.514 &   \\  
  258.3710   &  \six      &  6.1   & 2737.30 & 4175.65 & 3144.01 &   \\  
  261.0440   &  \six      &  6.1   & 1151.82 & 1731.62 & 1318.26 &   \\  
  262.9760   &  \fexvi    &  6.4   & 173.934 & 968.696 & 1036.80 &   \\   
  263.7657   &  \fexxiii  &  7.2   & 14.5618 & 12.5289 & 9.42194 &  u   \\ 
  264.2310   &  \sx       &  6.2   & 925.718 & 975.778 & 899.526 &   \\   
  264.7900   &  \fexiv    &  6.3   & 2784.31 & 5570.13 & 4768.17 &   \\   
  265.0010   &  \fexvi    &  6.4   & 19.6691 & 105.484 & 99.7671 &   \\    
  270.5220   &  \fexiv    &  6.3   & 1356.52 & 3060.05 & 2607.03 &   \\   
  272.0060   &  \six      &  6.1   & 1117.27 & 1592.69 & 1191.82 &   \\  
  272.6390   &  \sivii    &  5.8   & 134.686 & 237.159 & 99.4883 &   \\   
  275.3540   &  \sivii    &  5.8   & 456.285 & 706.011 & 317.491 &   \\   
  284.1630   &  \fexv     &  6.3   & 10317.8 & 28644.9 & 27965.3 &   \\   

\enddata 
\tablenotetext{a}{ ``u'' indicates upper limits of fluxes of hot lines
which are not actually detected in the \eis\ spectra. }
\end{deluxetable*}

In the following we will describe in detail the analysis methods and 
results obtained for one of the selected regions, EHH. Then we will discuss
our findings for all three regions.

\subsection{Thermal Structuring from \eis\ spectra \label{ss:eis}}

For the analysis of \eis\ spectra and \xrt\ data, we use the Package 
for Interactive Analysis of Line Emission (PINTofALE, \citealt{PoA}) 
which is available on SolarSoft. For the spectral analysis and 
emission measure distribution reconstruction from \eis\ spectra 
we use CHIANTI v.6.0.1 \citep{chianti,chianti6}, with the ionization 
balance of \cite{Bryans09}.
Unless explicitly stated otherwise we assume ``coronal'' plasma 
abundances of \citet{Feldman92}.

The observed line intensities provide constraints on the plasma 
temperature distribution, as they depend on the plasma emissivities
$G_{\lambda}(T,n)$, element abundances $A_Z$, and differential
emission measure distribution $DEM (T)$:
\begin{equation}
I_{\lambda} = A_Z \int_{T} G_{\lambda}(T,n_e) DEM(T) \,dT
\label{eq:Iline}
\end{equation}
where $ DEM(T) = n_e^2 \,dV / dT$ $[{\rm cm}^{-3} {\rm K}^{-1}]$. 

Several lines listed in Table~\ref{tab:eislines} are density sensitive,
so it is necessary to measure the electron density of the emitting
plasma for each region before calculating the contribution functions
to be used in the analysis. We have used the line intensity ratios of
\ion[Si x], \ion[Fe xii] and \ion[Fe xiv] available in the dataset,
and the results are listed in Table~\ref{tab:density}. The most reliable
indicator of the electron density is \ion[Si x], since its lines are
not affected by blends and have been found to be reliable by
\cite{Young98} and \cite{Landi02}. The \ion[Fe xiv] intensity ratio is
less sensitive to the electron density and provides a much coarser
estimate; the density values it indicates are broadly consistent with
the \ion[Si x] measurements.  The \ion[Fe xii] densities, while
consistent within the uncertainties, are a bit larger than those of
\ion[Si x], a behavior that has already been noted elsewhere
\citep{Young09,Watanabe09}; this might be due to atomic physics
problems affecting the \ion[Fe xii] emissivities (however, see also
\citealt{Tripathi10}). Throughout the present work, we used $\log n_e
= 9.0$ for all regions in the calculation of the spectral emissivities
from CHIANTI. 

\begin{deluxetable}{lcccc} 
 \small
\tablecolumns{5} 
\tablewidth{0pc} 
\tablecaption{Electron density diagnostic results for each region. The
  density adopted throughout the study is $\log n_e = 9.0$. Densities
  are in cm$^{-3}$.	\label{tab:density}} 
\tablehead{ 
\colhead{Ion}  & \colhead{Line ratio}  & \multicolumn{3}{c}{$\log n_e$} \\
\colhead{}     & \colhead{} & \colhead{EC} & \colhead{EH} & \colhead{EHH} }
\startdata 
 
\ion[Si x]   & 261.0/253.8  & 8.95$^{+0.30}_{-0.20}$  & 9.00$\pm$0.25       &  8.95$^{+0.30}_{-0.20}$  \\
\ion[Fe xii] & 186.8/195.1  & 9.15$^{+0.15}_{-0.20}$  & 9.20$\pm$0.20       &  9.10$\pm$0.15         \\
\ion[Fe xiv] & 264.8/270.5  & 9.5$\pm$0.4           & 9.3$^{+0.3}_{-0.5}$   &  9.3$^{+0.3}_{-0.5}$     \\

\enddata 
\end{deluxetable}

\begin{figure}[!ht]
\centerline{\includegraphics[scale=0.6]{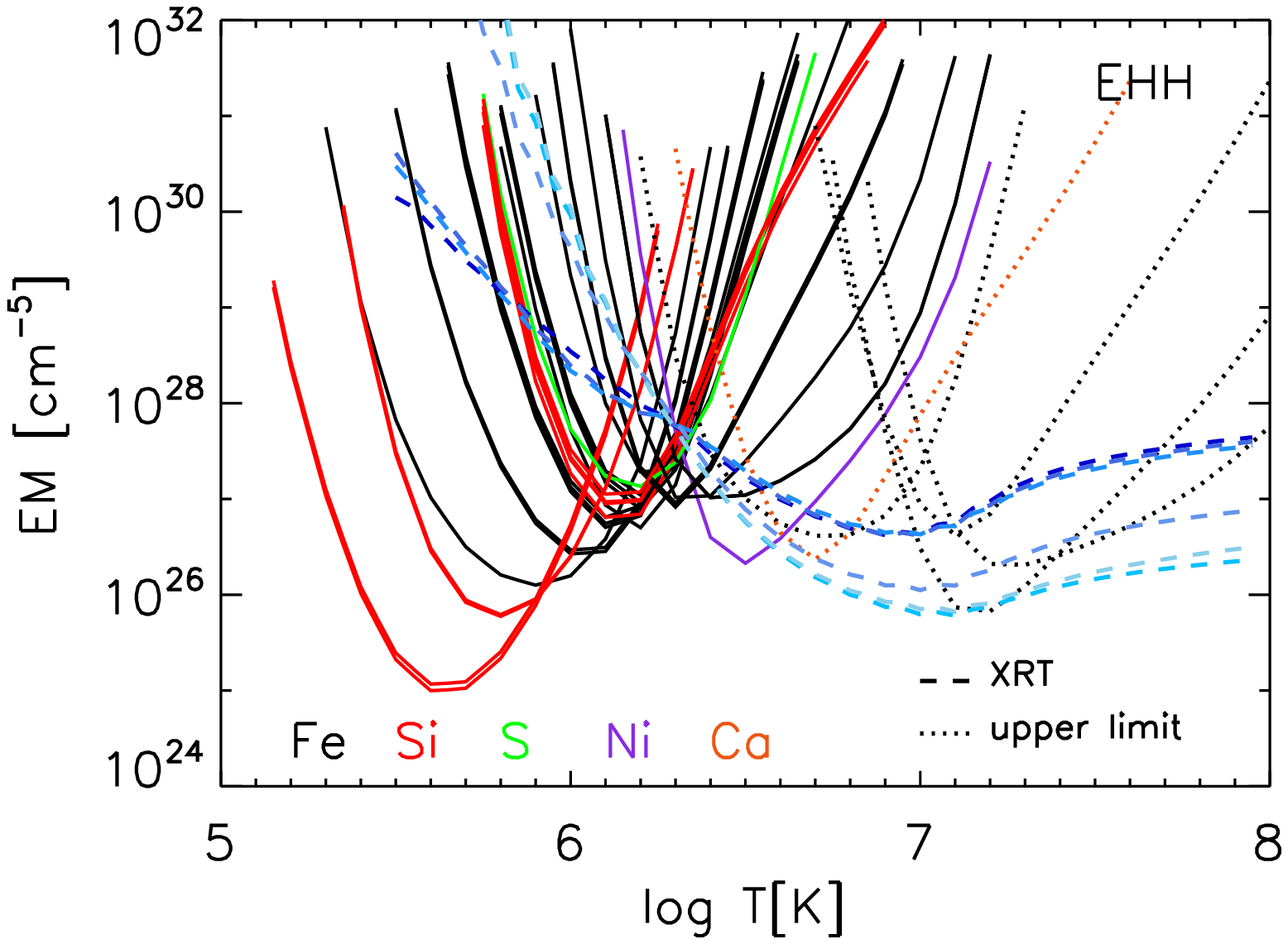}}
\caption{EM loci obtained for the measured \eis\ line fluxes, including 
	upper limits (dotted lines) for hot lines in \eis\ passbands. 
	The dashed lines indicate the EM loci for \xrt\ measured fluxes.
	We use different colors for lines of different elements,
	as indicated in the inset. 
	\label{fig:EMloci}}
\end{figure}

Figure~\ref{fig:EMloci} shows the EM loci corresponding to the fluxes 
measured for one of the regions. The curves represent the loci of constant
flux for a given line, i.e.\ they are obtained by taking the ratio of 
the measured line flux to the emissivity function (including the element
abundance) of that line. The EM loci plot provides a visual indication
of the temperature range covered by the selected lines and passbands, since
we are also including EM loci for the \xrt\ observations. For the XRT
channels the EM loci are derived by dividing the measured fluxes by
the temperature response of that channel, derived assuming a given set
of element abundances (in the case of Fig.~\ref{fig:EMloci} we use
coronal abundances). Also, if the 
underlying plasma were isothermal the EM loci curves would all cross at 
a single points, therefore this plot also provides some indication for 
the characteristics of temperature distribution of the plasma. 

The method of reconstruction of the differential emission measure 
distribution that we adopt for the analysis of 
\eis\ data runs a Markov-chain Monte Carlo (MCMC) algorithm on a 
set of line fluxes (intensities in broad bandpasses can also be 
used, as for instance in the case of \xrt\ datasets; see 
\S\ref{ss:xrt}), and it returns an estimate of the DEM that generates 
the observed fluxes (see \citealt{Kashyap98} for details on assumptions 
and approximations). 
With respect to other methods for reconstructing the plasma emission 
measure distribution from a set of fluxes, the MCMC method we adopt 
has the main advantage of providing an estimate of the uncertainties 
in the derived DEM. The problem of determining the emission measure
distribution and its confidence limits is notoriously challenging
\citep[e.g.,][]{Craig76,Judge97,Judge10}. 
Part of the reason is that the emission measure at a given temperature
cannot be determined independently of temperatures at other bins, and
therefore the corresponding errors are also correlated.  The MCMC
method works around this fundamental problem by sampling solutions
from the full probability distribution of the DEM given the data.  It
is a feature of the MCMC chain that regardless of the conceptual
complexity of the solution space, it can be fully explored numerically
at a relatively low computational cost.  Thus, the sampled solutions
include the effects of statistical noise from the measured data as
well as correlations across temperatures that arise due to overlaps
between the individual contribution functions of the different lines.
From the set of solutions thus obtained, we can depict the uncertainty
range at temperatures of interest.  Since it is difficult to display
correlations, for purposes of clarity, we only show the error ranges
computed separately at each temperature bin in the figures.  Error
bars computed for predicted fluxes, and thus abundances, are
uncorrelated and thus have the usual meaning.  Another problem with
DEM reconstruction is that the derived curves are solutions to a
Fredholm integral equation of the first kind, and are thus subject to
high-frequency instability.  These instabilities are typically
suppressed by imposing a global smoothness criterion.  In the case of
the MCMC method adopted here however, this restriction is relaxed.
Smoothing is more physically based, is locally variable, and is
limited by the width and number of the line contribution functions
used.  This leads to solutions that individually have more
fluctuations, but on average the ensemble of solutions produce an
envelope that reliably determines features that are statistically
significant. 

As uncertainties on the measured line fluxes listed in 
Table~\ref{tab:eislines} we combine in quadrature the statistical
uncertainties which are typically very small (of the order of 
few percent), with the uncertainty in the absolute radiometric
calibration for \eis\ which is estimated to be 22\% \citep{Lang06}.   

\begin{figure}[!ht]
\centerline{\psfig{figure=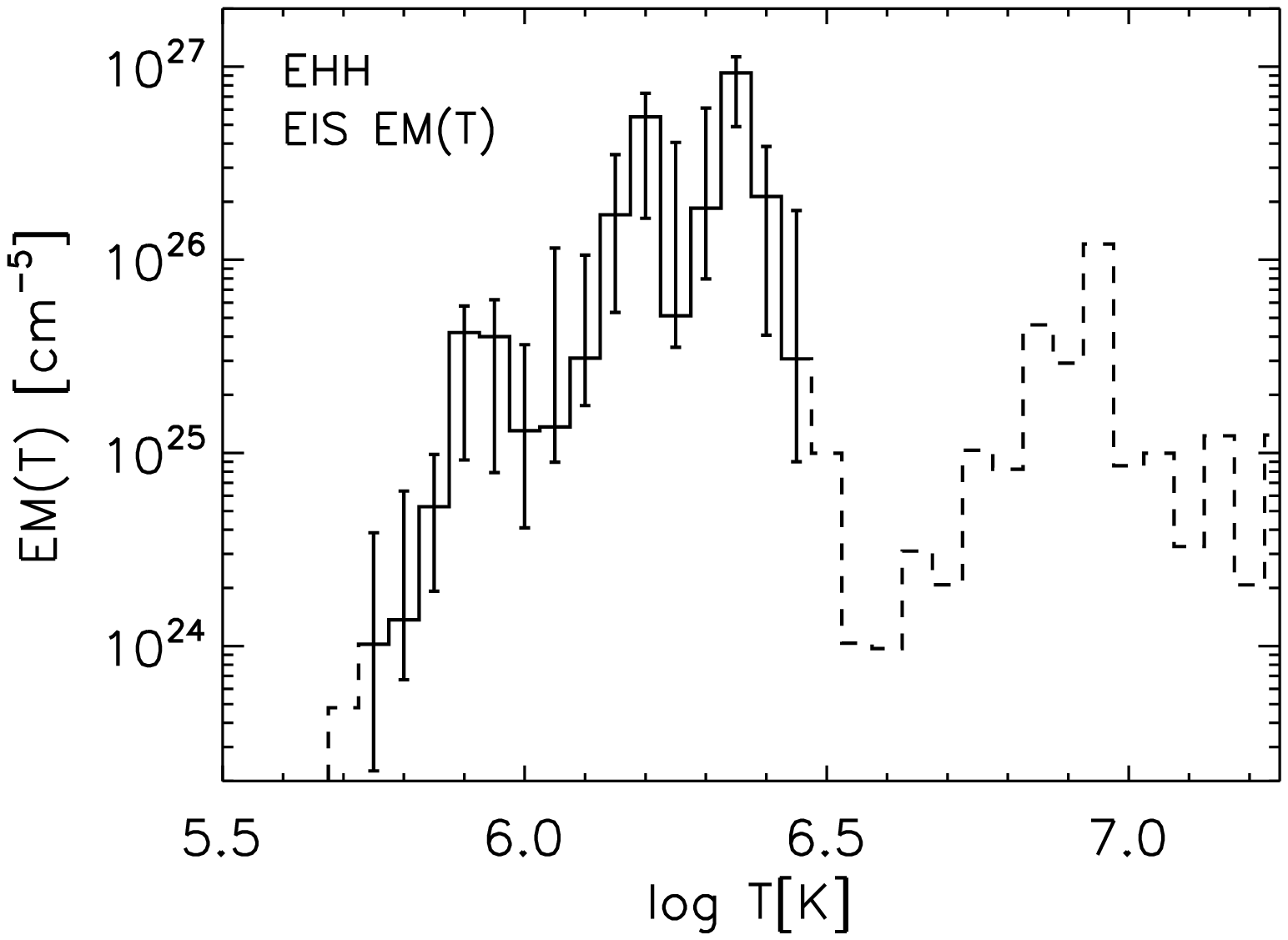,width=9cm}}
\centerline{\psfig{figure=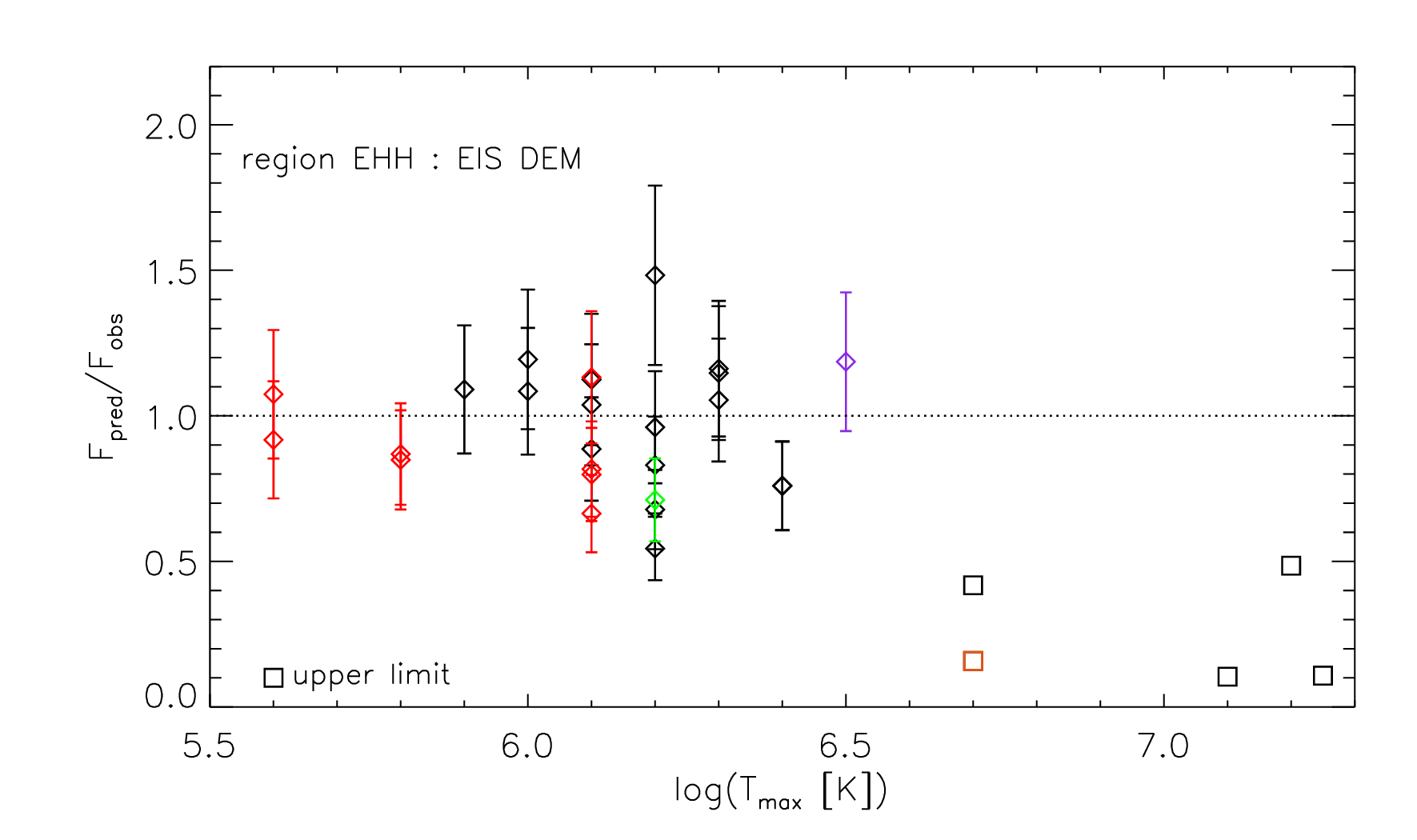,width=9cm,angle=90}}
\caption{{\em Top:} Emission measure distribution derived for region EHH 
	with the MCMC method (see text) using \eis\ line fluxes listed 
	in Table~\ref{tab:eislines}. 
	The error bars provide a measure of the extent to which the \emt\ of 
	each temperature bin is constrained by the measured fluxes.
	We plot uncertainties only for the temperature bins for which the
	\emt\ contributes to at least 5\% of the flux of at least one line 
	(see Figure~\ref{fig:flx_contrib}, and text for details).
        Unconstrained bins are plotted with dashed line.
	{\em Bottom:} Comparison of measured \eis\ line fluxes with flux 
	values predicted using the emission measure distribution derived 
	with the MCMC method. Different colors are used for lines of 
	different elements, in accordance with the colors used to show the 
	EM loci (Figure~\ref{fig:EMloci}).
	\label{fig:E_EM_PrOb}}
\end{figure}

Figure~\ref{fig:E_EM_PrOb} shows the results of the MCMC method to 
derive the emission measure distribution, applied to the \eis\ fluxes 
measured for the EHH region. In this case, as in the rest of the 
paper, we will plot the emission measure distribution \emt\, which 
is obtained by integrating the differential emission measure 
distribution $DEM(T)$ in each temperature bin (here we use a
temperature grid with constant $\Delta \log T = 0.05$).
The upper panel shows the emission measure 
distribution that reproduces the measured fluxes; the associated 
uncertainties estimated as described at the beginning of this section
are also plotted.
The bottom panel shows the ratios of predicted to observed fluxes 
for the \eis\ lines.
The comparison of measured fluxes with predictions based on the 
emission measure distribution shows that all fluxes are reproduced
within about 50\%, and that the fluxes predicted for the hot lines,
for which the spectra only provide upper limit, are all lower than
the upper limits.

\begin{figure}[!ht]
\centerline{\psfig{figure=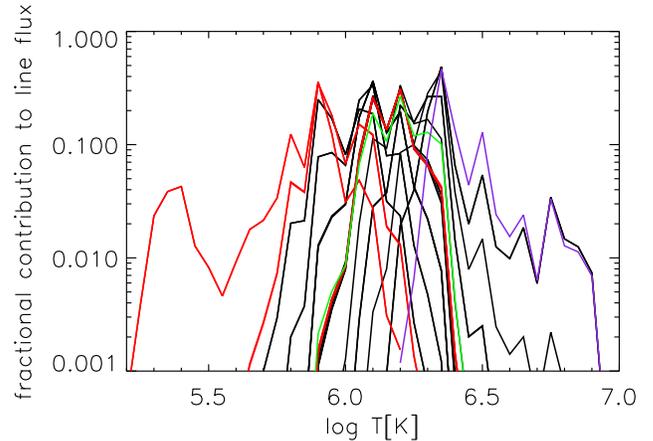,width=9cm}}
\caption{Each curve represents the fractional contribution of each temperature 
	bin to the flux of one of the \eis\ lines used for DEM reconstruction 
	(hot lines for which we only derived upper limits are not shown). 
	We use the same colors as in Figure~\ref{fig:EMloci}.
	\label{fig:flx_contrib}}
\end{figure}

We can investigate the compatibility of this derived \emt\ with the 
\xrt\ observations by computing the predicted \xrt\ fluxes and comparing
them with the measured ones. It is worth noting however that when 
deriving the \emt\ using exclusively the information contained in the
\eis\ spectra, not all temperature bins are well constrained. In particular
for these \eis\ datasets which lack measured lines in the high temperature
range ($\log T[{\rm K}] \gtrsim 6.5$) the hot tail of the \emt\ is poorly
constrained, but \xrt\ is very sensitive to it, having temperature responses
which peak around $\log T[{\rm K}] \sim 7$ for all filters. Therefore
some caution must be applied to perform this cross-check. 
To determine in which temperature bins the \emt\ is constrained by
\eis\ data we look at the contribution of each bin to the line fluxes.
Folding the \emt\ with the line emissivity we can investigate the
relative weight of each temperature bin for each line, as shown in 
Figure~\ref{fig:flx_contrib}.  We then consider as constrained the 
temperature bins where the \emt\ contributes more than a threshold 
(percentage) value to at least one of the \eis\ lines. 
The choice of the threshold value is somewhat arbitrary and we chose a 
rather conservative 5\%.

Considering the \emt\ in the temperature bins satisfying our selection 
criterium, we calculate the predicted fluxes in the \xrt\ filters 
(which therefore are
strictly speaking {\em lower limits} to the values predicted by the 
assumed \emt) which are listed in Table~\ref{tab:EHH_EisDem_XFlx}, 
together with the measured fluxes.
The comparison of the two sets indicates that the \emt\ derived from 
\eis\ underpredicts the \xrt\ fluxes by a factor $\sim 2$, the discrepancy 
being slightly larger for the thinner filters (Al-poly, C-poly, Ti-poly) which 
are more sensitive to the cooler plasma which is better characterized 
by the \eis\ data. 

\begin{deluxetable}{cccc} 
 \small
\tablecolumns{4} 
\tablewidth{0pc} 
\tablecaption{For region EHH, comparison of \xrt\ measured fluxes with
  the values predicted using the \emt\ derived from \eis\ line
  fluxes (and shown in the upper panel of Figure~\ref{fig:E_EM_PrOb}),
  and from the forward modeling with two temperature components (see
  \S\ref{ss:xrt}), assuming coronal  abundances.
	\label{tab:EHH_EisDem_XFlx}} 
\tablehead{ 
\colhead{\xrt\ filter}  & \colhead{F$_{\rm obs}$}  & 
	\colhead{F$_{\rm pred, EIS}$\tablenotemark{a}}  &
	\colhead{F$_{\rm pred, fm}$\tablenotemark{b}} \\
\colhead{}  & \colhead{[DN/s]}  & \colhead{[DN/s]}  & \colhead{[DN/s]}  }
\startdata 
 
Al-poly  &  122.0     &  57.8  (41.8-70.2)    &  123.2 \\ 
C-poly   &   95.8     &  44.5  (32.8-53.4)    &  101.1 \\
Ti-poly  &   69.1     &  31.0  (23.0-36.9)    &  65.94 \\
Be-thin  &   11.1     &  5.85  (3.69-7.61)    &  9.82 \\
Be-med   &    1.59    &  1.06  (0.66-1.38)    &  1.42  \\
Al-med   &    0.730   &  0.483 (0.301-0.628)  &  0.673 

\enddata 
\tablenotetext{a}{\xrt\ fluxes predicted using the emission
	measure distribution derived using \eis\ line
	fluxes only. The values in parentheses represent
	the range of fluxes predicted by the Monte Carlo
	simulations of the EM(T) which are the acceptable
	emission measure distributions, defining
	the error bars shown in Figure~\ref{fig:E_EM_PrOb}. }
\tablenotetext{b}{\xrt\ fluxes predicted using the \emt\
	derived from the forward modeling with 2T, and using 
	coronal abundances.}
\end{deluxetable}

\subsection{Thermal Structuring from \xrt\ data \label{ss:xrt}}

We then derive the thermal distribution of the coronal plasma 
exclusively from \xrt\ data, by using two different methods:
\begin{itemize}
\item Forward fitting of distributions of filter ratios of several 
	filter pairs, through pixel-by-pixel Monte Carlo simulations 
	of the observations, using two temperature components (2T);
\item MCMC method, as for the analysis of the \eis\ spectra (see 
	\S\ref{ss:eis}).
\end{itemize}
We use \xrt\ temperature responses calculated assuming ``coronal''
abundances \citep{Feldman92} and the new \xrt\ calibration which
includes a model for the contamination layers deposited both 
on the CCD and on the focal plane filters (N.Narukage et al.,
2010, Solar Physics, submitted).
For the detailed comparison of \xrt\ and \eis\ data for each of
the selected region we only use the subset of \xrt\ data taken
simultaneously to the \eis\ observations (see Fig.~\ref{fig:lc_xrt}).

We have applied the forward fitting method with two T components
already in previous analysis
of \xrt\ data aimed at deriving the plasma temperature distribution of 
an active region, in particular to investigate the presence of a high 
temperature component \citep{Reale09}. The details of this method are 
thoroughly discussed by \cite{Reale09}. 
In summary, the pixel fluxes in the relevant \xrt\ filters are
computed from a basic model EM(T) along the line of sight defined by
few parameters (amplitude, central temperature, and width of either
single or double top-hat functions). Monte Carlo simulations are
used to randomize some of the EM(T) parameters and to include
Poisson photon noise. This procedure is replicated to build whole
fake \xrt\ images in all filters. The simulated \xrt\ data are then
analyzed exactly as they were actual data: (1) temperature and
emission measure maps, i.e., single values for each pixel, are
obtained for a given filter ratio, and this is done for several
filter pairs which are sampling the thermal structure in different
ways; (2) for each considered filter ratio an emission measure
distribution vs.\ temperature is built by summing up the emission
measure values of all pixels falling in the same temperature bin.
The input \emt\ is deemed an adequate representation of the 
underlying plasma emission measure distribution when these curves of 
emission measure distributions vs.\ temperature are reproducing 
satisfactorily the analogous curves obtained from actual data.

\begin{figure}[!ht]  
\centerline{\includegraphics[scale=0.45]{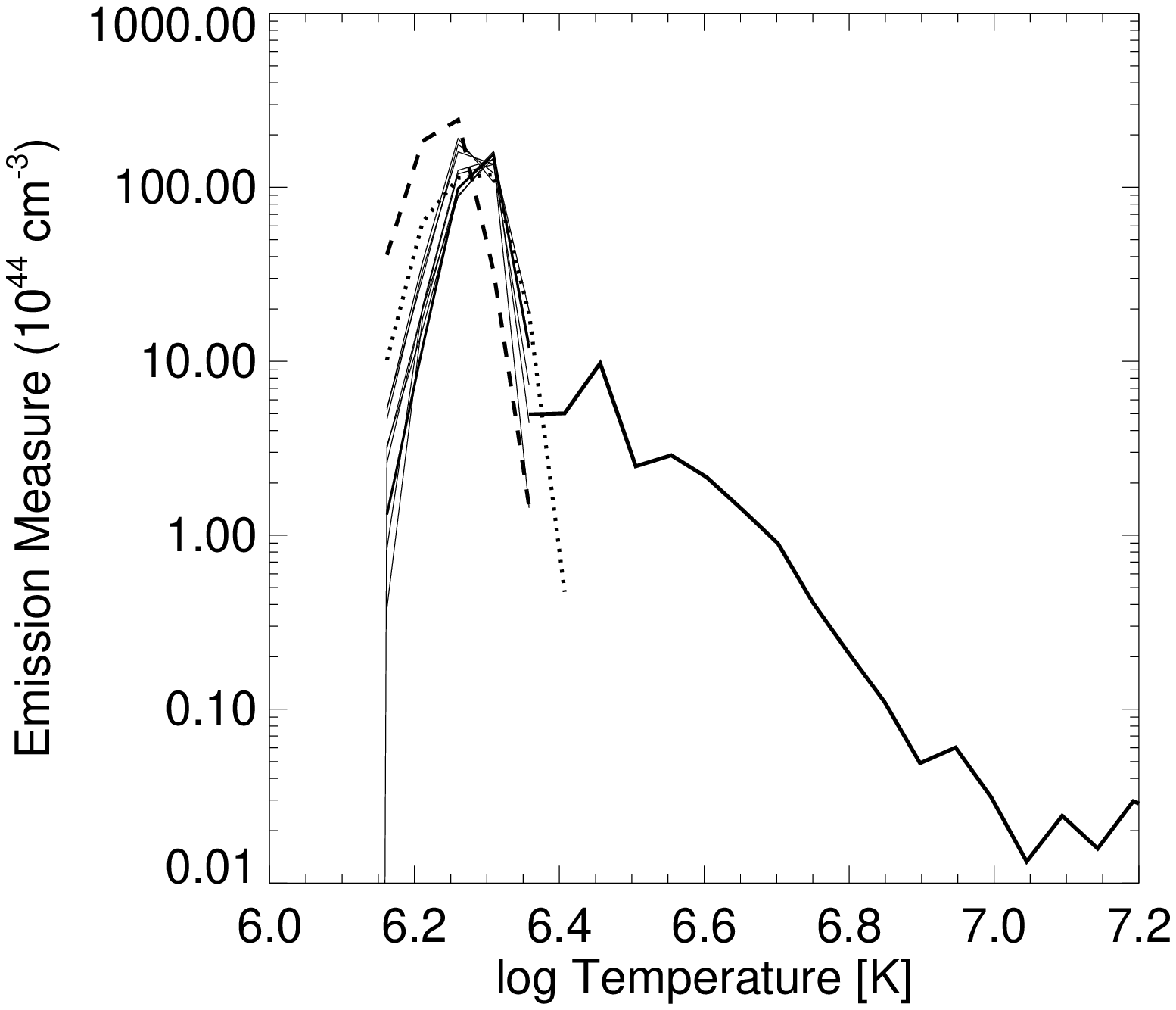}} \vspace{-0.7cm}
\centerline{\includegraphics[scale=0.45]{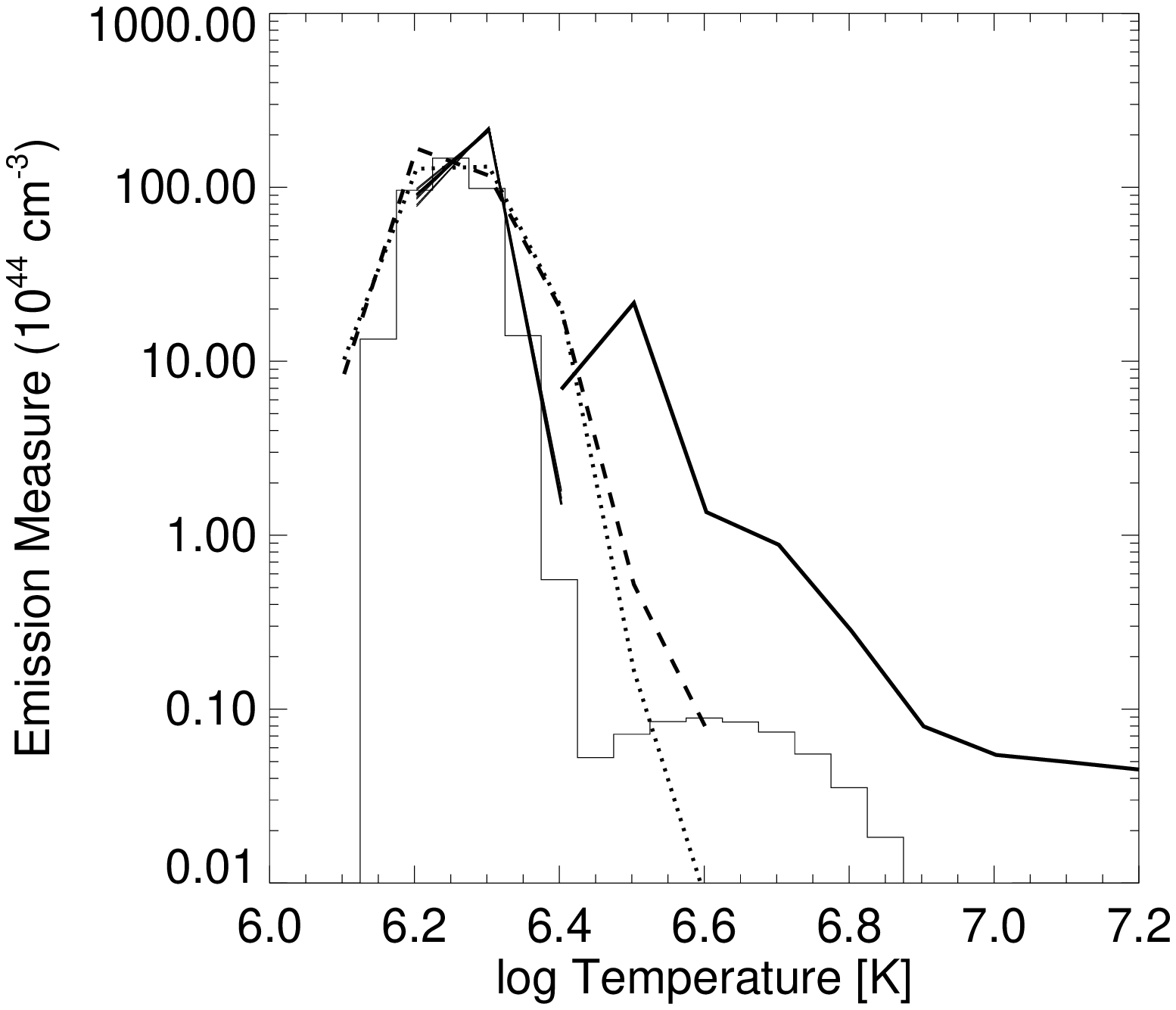}}
\caption{{\em Top:} Emission measure distribution vs.\ temperature
	derived for region EHH from \xrt\ data. Each curve correspond
	to a filter pair: soft filter ratios (thin solid lines), 
        Al\_med/Be\_thin (dotted line), Be\_med/Be\_thin (dashed
        line), Be\_med/Al\_med (thick solid line).
	{\em Bottom:} Emission measure distribution vs.\ temperature
	derived for region EHH from \xrt\ data simulated using the 
	\emt\ histogram as parent emission measure distribution 
	along the line of sight. Lines as in the top panel.
	\label{fig:sims_xrt}}
\end{figure}

The results of these simulations for region EHH are shown in 
Figure~\ref{fig:sims_xrt}, and Table~\ref{tab:EHH_EisDem_XFlx}. 
We note that, as discussed thoroughly 
in \cite{Reale09} for the analysis of \xrt\ observations of another
active region, while this analysis method is certainly approximate
it allows us to be sensitive to possible minor EM(T) components,
e.g., small hot components, and to somewhat constrain them. 
As for the case presented in \cite{Reale09}, also here we find that a
high temperature plasma component, even if much weaker than the
dominating cool ($\log T[{\rm K}] \sim 6.1-6.3$), is nevertheless
appropriate to reproduce the \xrt\ observations.

\begin{figure}[!ht]
\centerline{\psfig{figure=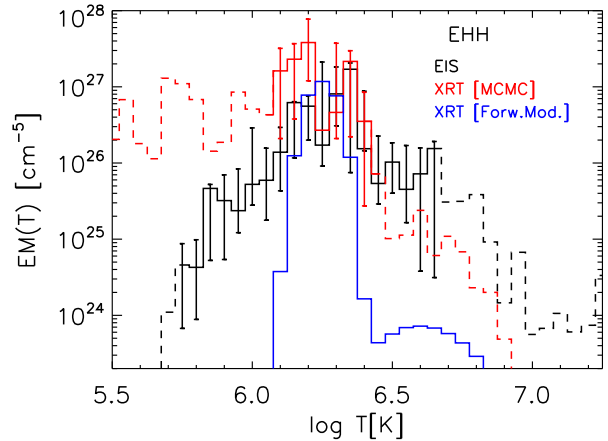,width=9cm}}
\caption{Comparison of emission measure distribution derived for region 
	EHH independently from \eis\ lines (with MCMC method, see 
	\S\ref{ss:eis}; black line), and from \xrt, with two different 
	methods: MCMC (red line), and forward modeling with two
        temperature components (blue line). 
	For the \emt\ derived using the MCMC method we show the 
	estimate of uncertainties for those temperature bins which 
	contribute to at least 5\% to the flux of at least one line 
	or passband (see Figure~\ref{fig:flx_contrib}).
	\label{fig:EM_E_X}}
\end{figure}

As mentioned above, in order to derive the \emt\ from \xrt\ data we
also use the MCMC method, using the flux values integrated in each
passband over the whole subregion. 
{ The errors associated to the \xrt\ fluxes are estimated by taking
into account the photon noise and the calibration uncertainties. 
Like for the \eis\ fluxes, the statistical uncertainties (we observe
bright sources and integrate over time to further increase the
signal-to-noise ratio) are typically very small and the errors are
largely dominated by the calibration uncertainties (for an estimate of
the latter see Narukage et al.\ 2010, Solar Physics, submitted). The
resulting errors are of the order of $\sim 5$\% for the thin filters,
and $\sim 15-20$\% for thin-Be and the medium filters.}
The derived best fit \emt\ is shown in Figure~\ref{fig:EM_E_X} (red
line); the associated uncertainties are rather large, as should also
be expected considering the limited constraints that the \xrt\ fluxes
in six filters can provide to the emission measure distribution over a
large temperature range.

The differences between the results of the two methods are not
  large, compared to the associated uncertainties.
The agreement is overall acceptable if one considers that the
approaches are completely different. The MCMC method is model
independent but it only uses the total region flux values. 
The forward fitting 2T method makes an assumption on the basic EM(T)
distribution but it keeps the spatial information of the pixel values
longer in the analysis pipeline, and it is therefore more sensitive to
small local plasma inhomogeneities (at the scale of the single
pixel). Moreover this latter method gives large weight to the
Be\_med/Al\_med filter ratio, and this allows to detect the small hot
component.

In Figure~\ref{fig:EM_E_X} we compare the emission measure 
distributions derived independently from \xrt\ and \eis\ data.
Considering only the temperature regions where the EM(T) are
constrained by the data, i.e.\ the bins where the error bars
are shown, the different datasets yield emission measure 
distributions that are not too dissimilar when the (large) 
uncertainties are taken into account. In particular,
they have similar peak temperature and width of the cool component. 
Using the \emt\ derived from \xrt\ data with the MCMC method 
to calculate the expected 
fluxes for the \eis\ lines of Table~\ref{tab:eislines}, we find that
the \xrt\ \emt\ reproduces the measured \eis\ fluxes not too
accurately. 
Specifically, the lines with typical temperature of formation 
$\log T[{\rm K}] \gtrsim 6.3$, such as \fexiv\ and \fexv, and the 
lines formed around $\log T[{\rm K}] \sim 6.1-6.2$, such as \fexii\ 
and \six, are overpredicted by a factor $\sim 3$ (typical range spans
approximately from a factor 2 to 5); cooler lines 
($\log T[{\rm K}] \lesssim 6.0$) are instead slightly underpredicted
(typical range: $\sim 0.05$ to $\sim 1.5 \times$ the measured fluxes).

\subsection{Thermal Structuring from \eis\ and \xrt\ \label{ss:eisxrt}}

\begin{figure}[!ht]
\centerline{\psfig{figure=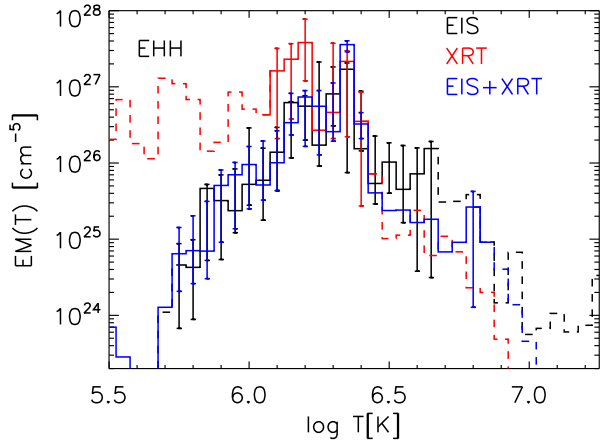,width=9cm}}
\caption{Comparison of emission measure distribution derived for region 
	EHH with different methods: from \eis\ lines (black line;
	same curve as Figure~\ref{fig:EM_E_X}), from \xrt\ (red
	line; same curve as curve shown in Figure~\ref{fig:EM_E_X}), 
	and using \eis\ and \xrt\ data together (blue curve).
	\label{fig:EM_E_X_XE}}
\end{figure}

We finally take advantage of all the available information, and derive
the plasma \emt\ by applying the MCMC method to the measured \xrt\ and 
\eis\ fluxes together.
The \emt\ derived for region EHH by combining the two instruments is shown in
Figure~\ref{fig:EM_E_X_XE}, compared to the emission measure 
distributions derived independently from each of the two 
datasets. 
When combining \xrt\ and \eis\ data, the resulting \emt\ is overall
rather similar to the \emt\ derived from \eis\ data, which provide
much tighter constraints because of the large number of lines, and
temperature dependence of their emissivity.
In the high temperature range however, where \xrt\ is more sensitive
to the plasma emission and \eis\ provides poor constraints, the 
\xrt+\eis\ \emt\ departs more from the \eis\ curve.

\begin{figure}[!ht]
\centerline{\psfig{figure=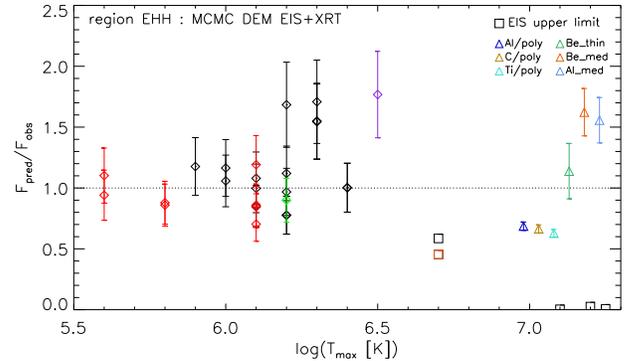,width=9cm,angle=90}}
\caption{Comparison of measured \eis\ and \xrt\ fluxes with flux 
	values predicted using the emission measure distribution derived 
	with the MCMC method using both set of fluxes and shown in
	Figure~\ref{fig:EM_E_X_XE}. We use the same colors as in 
	Figure~\ref{fig:EMloci} to indicate lines of different elements
	or \xrt\ passbands.
	\label{fig:XE_PrOb}}
\end{figure}

Figure~\ref{fig:XE_PrOb} shows how well the \xrt+\eis\ \emt\ reproduces
the measured fluxes. While the \eis\ line fluxes are reproduced 
reasonably well, similarly to the case of the \eis\ \emt, there is 
some systematic discrepancy between the predicted and the observed 
fluxes in the \xrt\  passbands. Specifically, the three cooler filters,
Al-poly, C-poly, Ti-poly, are underpredicted by about a factor 2.
The fluxes in the hotter filters are instead well reproduced. We note
that for these latter filters, which are much less sensitive to cool plasma
with respect to the thin filters, the algorithm of derivation of
the emission measure distribution can find a satisfactory agreement 
with the observations because of the leeway in the determination
of the high temperature component which is not tightly constrained
by the other fluxes, and in particular does not contribute
significantly to the analyzed EIS lines.
These discrepancies are reminiscent of the systematics found when
predicting \xrt\ fluxes using the \eis\ \emt\ (see 
Table~\ref{tab:EHH_EisDem_XFlx}), and when using the \xrt\ \emt\ to 
predict the \eis\ line fluxes (see discussion at end of \S\ref{ss:xrt}).
The cross-check provided by the three methods points to a systematic
difference with \eis\ data yielding emission measure values a factor
$\sim 2$ lower than the values derived from \xrt\ data, as also
evident from Figure~\ref{fig:EM_E_X_XE}.

We explored how some of the assumptions we have made for our analysis
may affect these results. Among those, the assumption for the element 
abundances of the coronal plasma has a potentially very significant 
effect on the derivation of the plasma emission measure distribution.
There is a large body of evidence that element abundances in the solar
corona do not reflect the solar photospheric composition
\citep[e.g.,][]{Meyer85,Feldman92}.  
In particular, the chemical fractionation appears to be a function of 
the First Ionization Potential (FIP), with low FIP elements, such as 
Fe and Si, typically enhanced in the corona by a factor of a few, 
whereas high FIP elements such as O are thought to have coronal 
abundances close to their photospheric values 
\citep[e.g.,][]{Meyer85,Feldman92}.  

The intensity of spectral lines is linearly dependent on the element 
abundance (assuming that the abundance is the same at all temperatures 
for plasma in the studied region; see Eq.~\ref{eq:Iline}). 
Therefore, if the \emt\ is derived from lines all emitted from a 
single element, e.g.\ Fe, a change in the element abundance will not 
change the shape of the emission measure distribution but merely shift 
its absolute value: for instance, a decrease of the abundance by a 
factor 2 will determine an increase of the \emt\ by a factor 2.
Although our analysis of \eis\ data includes lines emitted by different 
elements, the large majority of the considered lines are emitted by low
FIP elements Fe and Si. Therefore, if we consider a chemical composition
with a FIP bias different from the \cite{Feldman92} set, typically adopted 
for coronal plasma and used for all the above analysis, we expect the
shape of the \emt\ to not change by much qualitatively even its absolute 
values can change significantly.

\begin{figure}[!ht]
\centerline{\psfig{figure=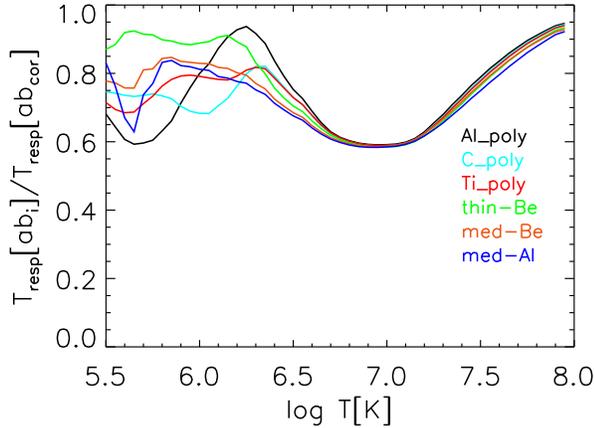,width=9cm}}
\caption{Ratio of \xrt\ responses, as a function of temperature, calculated 
	using intermediate abundances (FIP bias = 2), to the responses calculated using 
	coronal abundances \citep{Feldman92}. A different color is used 
	for each filter used in this work, as indicated in the inset. 
	\label{fig:XRTresp_abund}}
\end{figure}

For wide band instruments like \xrt\ the effect the adopted chemical
composition of the plasma on the temperature response is not as 
straightforward, as at different temperatures the relative contribution 
of the emission from different elements can change significantly.
In Figure~\ref{fig:XRTresp_abund} we show the effect of the element
abundances on the temperature response of the \xrt\ filters used in this
work. We consider the \cite{Feldman92} ``coronal'' abundances, and a 
set of abundances intermediate between coronal and photospheric 
\citet{GrevesseSauval}, i.e.\ with a FIP bias of 2 instead of 4 which
is typically assumed for coronal plasma.
For instance for this set of abundances Fe, Si and Mg have abundances
about half the values of \cite{Feldman92}.
Figure~\ref{fig:XRTresp_abund} shows that in the range 
$\log T[{\rm K}] \sim 6.7-7.3$ the change in the responses is approximately
linear with the abundances of these low FIP elements which are dominating
the emission in this high temperature range. 
At lower temperatures however the \xrt\ filters are affected differently 
depending on the details of their response as a function of wavelength 
and therefore the importance of different lines. 

\begin{figure}[!ht]
\centerline{\psfig{figure=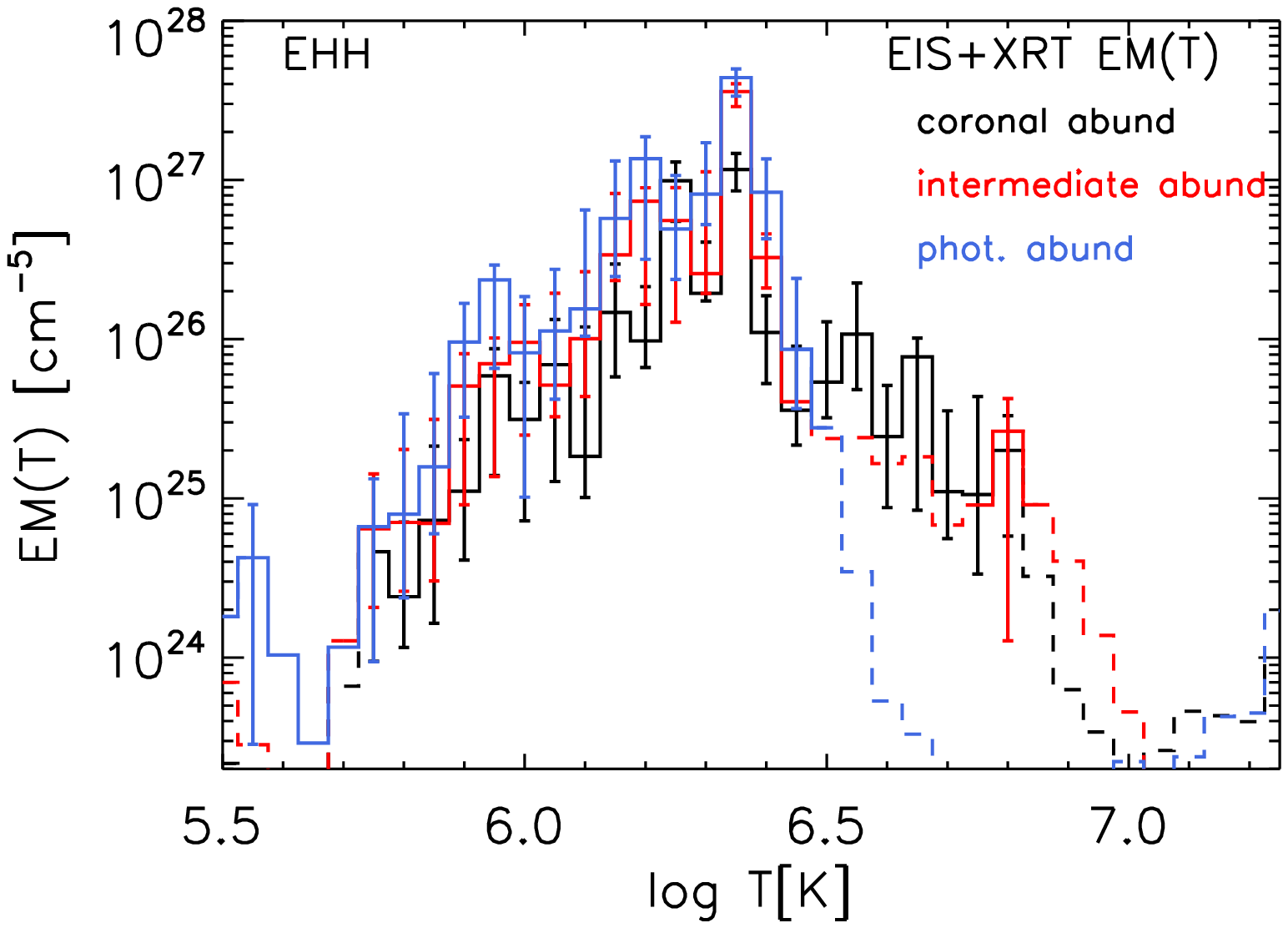,width=9cm}}
\centerline{\psfig{figure=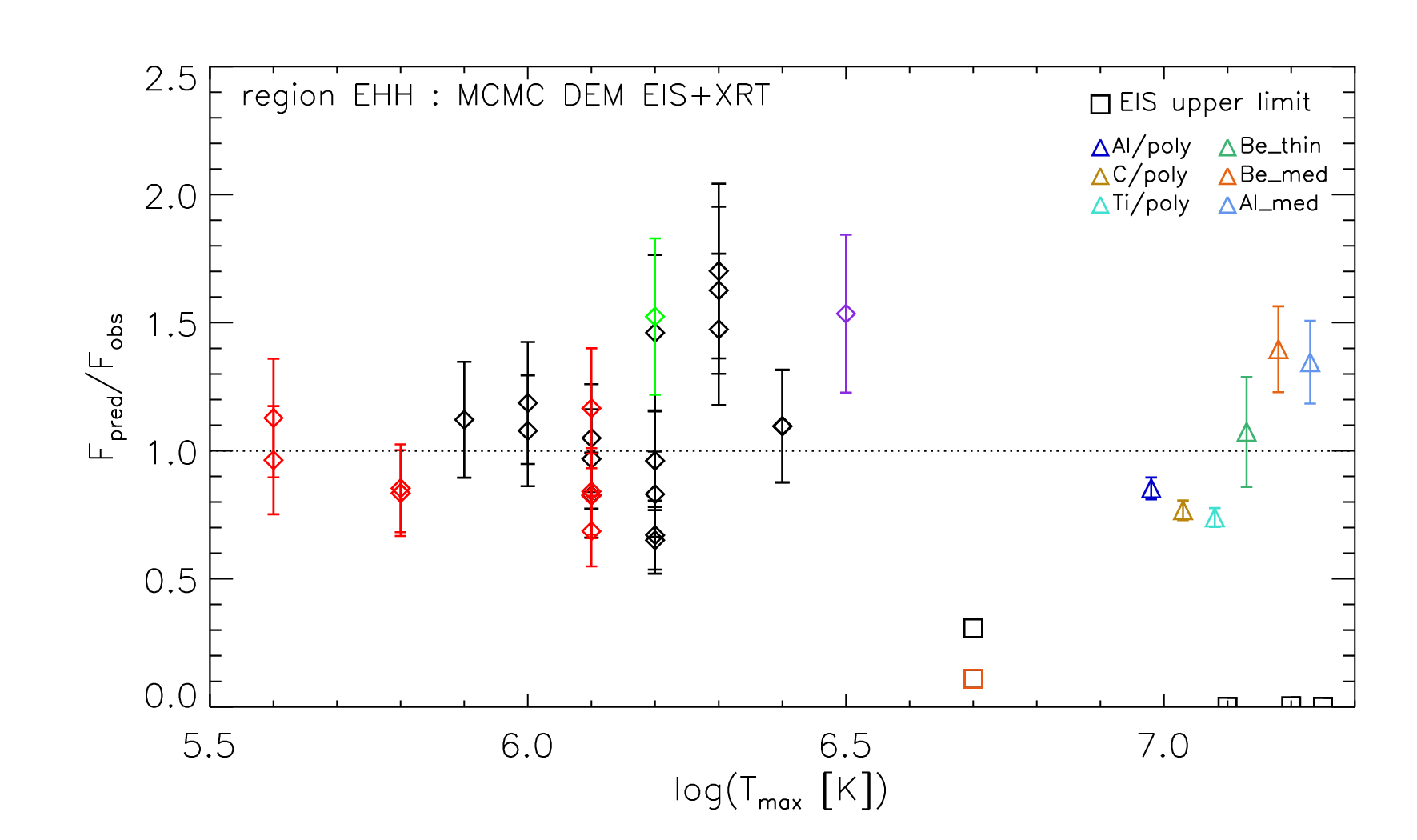,width=9cm,angle=90}}
\centerline{\psfig{figure=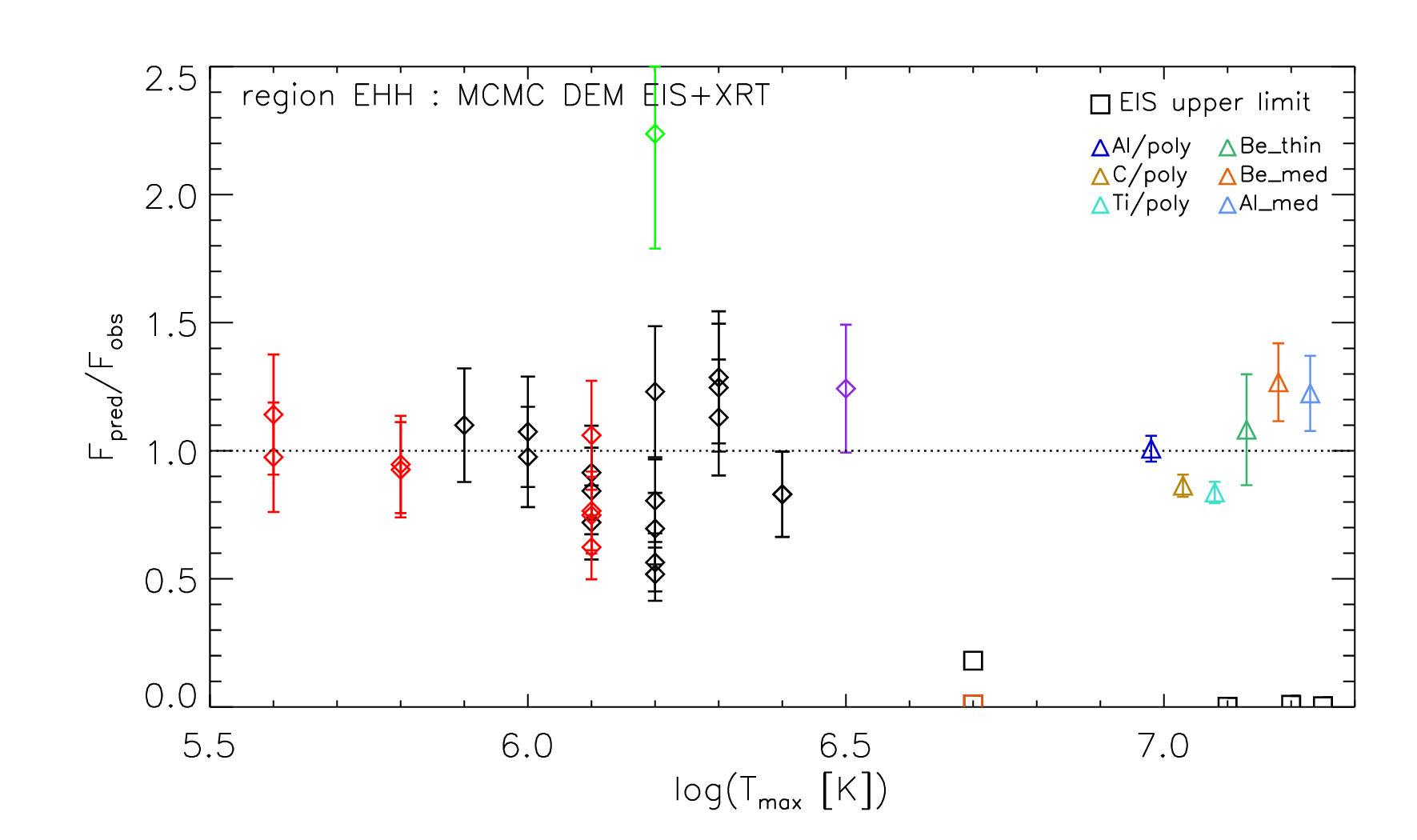,width=9cm,angle=90}}
\caption{{\em Top panel:} Comparison of emission measure distributions derived 
	with the MCMC method using simultaneously \eis\ and \xrt\ fluxes, 
	and three different sets of abundances: coronal abundances 
	(\citealt{Feldman92}; black line), intermediate abundances 
	(red line; see text for definition of intermediate
        abundances), and photospheric abundances
        (\citealt{GrevesseSauval}; blue line).
	For intermediate abundances, for which Fe and Si have roughly half
	the abundance as in the coronal set, the \emt\ accordingly
        increases; the \emt\ increases even further for photospheric abundances.
	{\em Middle and bottom panels:} Comparison of measured \eis\
        and \xrt\ fluxes with flux values predicted by the emission
        measure distribution derived with the MCMC method (\xrt+\eis)
        and: (a) intermediate abundances ({\em middle}), or (b)
        photospheric abundances ({\em bottom}; \citealt{GrevesseSauval}). 
	\label{fig:XE_EM_PrOb_abi}}
\end{figure}

We repeated our analysis using different sets of abundances, and
in particular we derived the \emt\ using this set of intermediate
abundances described here above, and the set of photospheric
abundances of \cite{GrevesseSauval}. The results we obtained are 
shown in Figure~\ref{fig:XE_EM_PrOb_abi}, also compared 
to the \emt\ derived with \citeauthor{Feldman92} abundances. 
The curves have very similar shape and are overall compatible within the 
uncertainties. However it is clear that, as expected in the 
light of the above discussions, the \emt\ obtained for intermediate
abundances (which are lower than the ``coronal'' abundances) is systematically
larger than the one found for ``coronal'' abundances, and the \emt\
derived for the photospheric abundances case are even larger.
In the middle panel of Figure~\ref{fig:XE_EM_PrOb_abi} we show the 
comparison of the observed fluxes with the predictions using the \emt\
derived for intermediate abundances.
Comparing these findings with the analogous results of Figure~\ref{fig:XE_PrOb}
we note that although the overall results are not changed dramatically,
the fluxes predicted for the \xrt\ thin filters are closer to the observations
when using intermediate abundances. The increases by about 20-35\% are 
compatible with expectations: the predicted fluxes should rise 
accordingly to the increase of the cool component of the \emt\, but 
be reduced by a factor corresponding to the decrease of the \xrt\ response 
(see Figure~\ref{fig:XRTresp_abund}).
The analogous plot for the case of photospheric abundances
(bottom panel of Figure~\ref{fig:XE_EM_PrOb_abi}) shows that using
these abundance values the agreement between \eis\ and \xrt\
improves even further. 

The values assumed for the element abundances can in 
principle be checked with the \eis\ spectra. As discussed above, the 
DEM obtained from \eis\ has been determined using almost exclusively 
lines from low-FIP  ions, whose coronal abundances are enhanced by 
a factor of $\sim 3-4$ over the photospheric values. 
Therefore, any change in the actual abundance of the low-FIP elements 
from the assumed value will result in a systematic shift between the 
observed intensities of the lines from \sx\ to \sxiii, whose 
coronal abundance is expected to be close to the photospheric one, 
and their values predicted with the DEM.
There are ten S lines in the \eis\ spectra, emitted by \sx\ to \sxiii, 
and we have compared their predicted and observed intensities to check 
the corrections to the coronal abundances. 
We found no unambiguous evidence of systematic differences in the 
observed to predicted intensity ratios. Moreover, recent estimates
of the S absolute abundance have been revised by \citet{Lodders03} and 
by \citet{Caffau07},  who decreased them by a factor 1.5 from 
the \citet{GrevesseSauval} value.
Since this change is of the same order of the revision of the low-FIP 
element abundances we propose, we feel that the uncertainties in the 
S predicted line intensities are so large that these ions can not 
be used to confirm the abundance change made necessary by the XRT 
channels in this work. 
S is in any case a borderline element for FIP effect studies, and
abundances of other low FIP elements are difficult to determine. 
While oxygen lines of O\,{\sc iv,v,vi} are present in the \eis\ wavelength
range, they are formed at low temperature ($\log (T[K]) \lesssim 5.5$)
with respect to the hotter coronal EIS lines considered in our active
region study. In that temperature range the thermal distribution is
not well constrained and therefore the abundances cannot be accurately
determined.

While the results shown in Figure~\ref{fig:XE_EM_PrOb_abi} seem
to suggest that the photospheric abundances might be a more
appropriate choice for the active region we studied here, we note
that the data we have used do not allow us to determine the
abundances and therefore do not allow us to disentangle between the
abundance effects and cross-calibration issues. 
We also note that this active region was already several days old at the
time of the observation, and therefore it is reasonable to expect some
significant effect of chemical fractionation to have occurred
producing departures from the photospheric composition (see e.g.,
\citealt{Widing01}; however, see also \citealt{DelZanna03}).

We have presented the complete analysis carried out for one of the
selected regions, providing insights into the thermal distribution
of the plasma, limitations of the methods, and cross-calibration of
the \hinode\ \xrt\ and \eis. Before discussing in details these findings
in \S\ref{s:discuss} we briefly present and discuss the analogous
results found for the other two selected regions, EC and EH, comparing
them with the results obtained for region EHH.

Carrying out the same analysis described above (\S\ref{ss:eis}, 
\ref{ss:xrt}, \ref{ss:eisxrt})
on the other two regions, we find results qualitatively similar to
our findings for region EHH discussed above, in particular in terms
of the comparison between different analysis methods and systematic
discrepancies between \xrt\ and \eis.

\begin{figure}[!ht]
\centerline{\psfig{figure=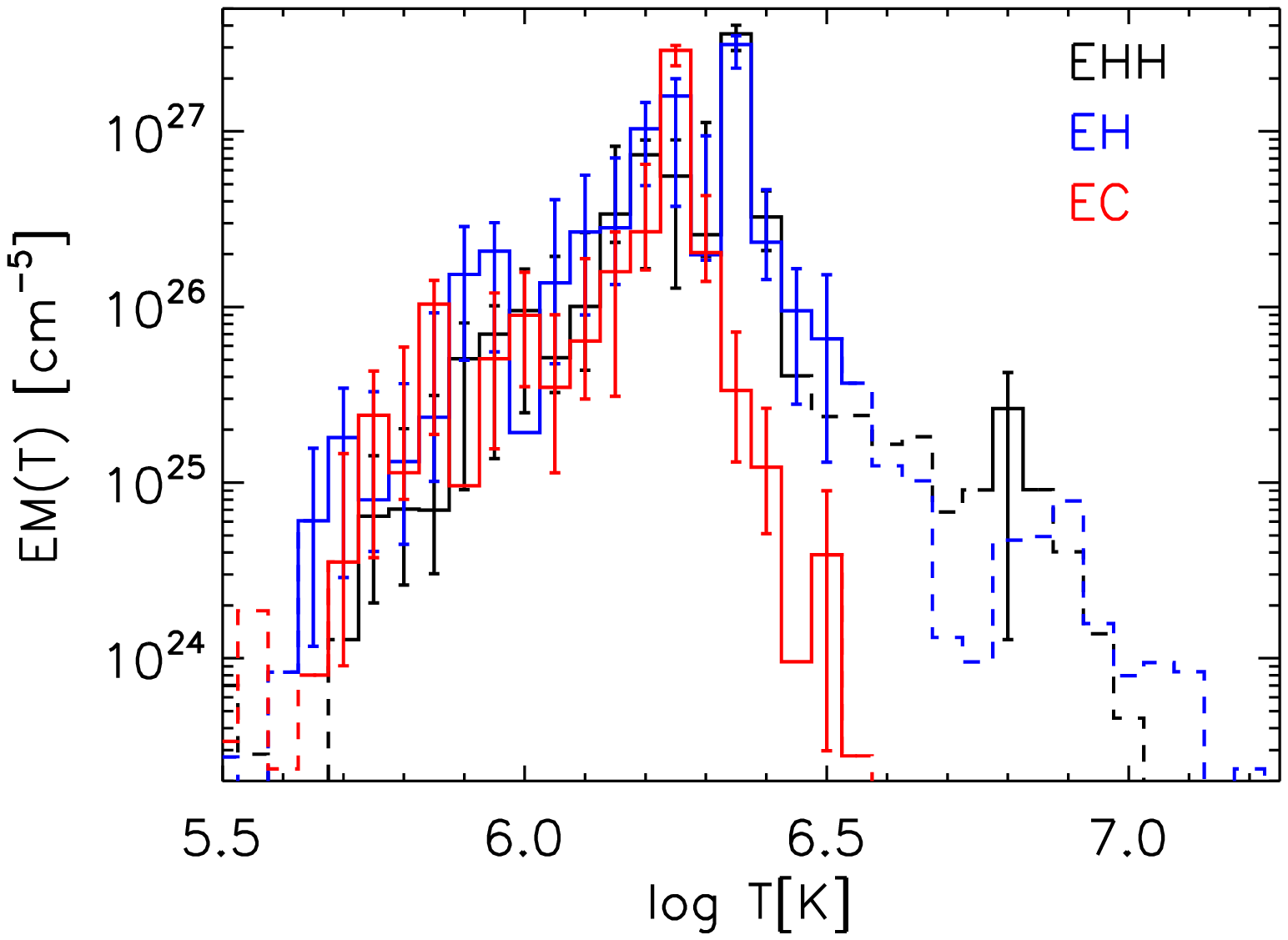,width=9cm}}
\centerline{\psfig{figure=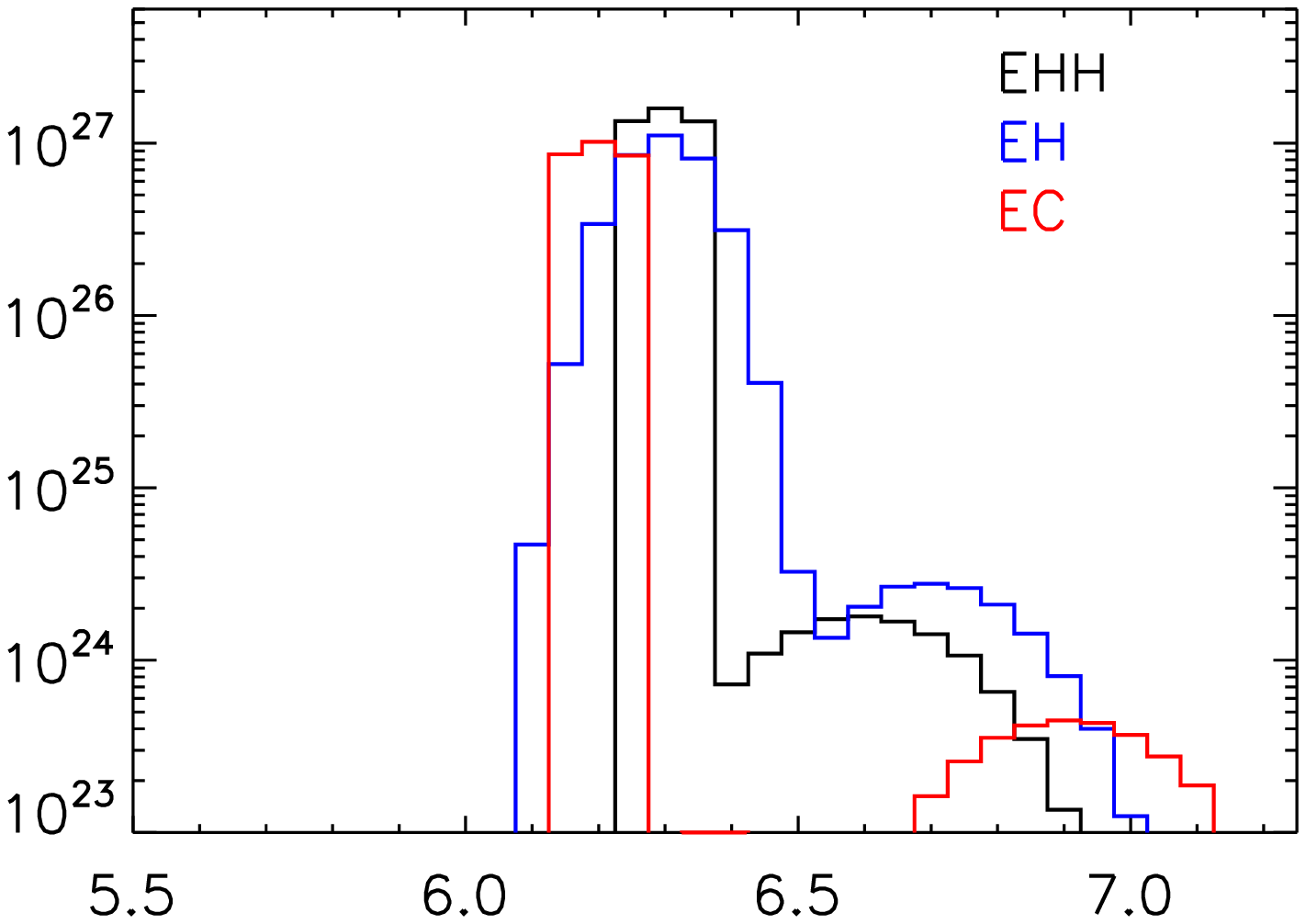,width=9cm}}
\caption{Emission measure distributions for the three selected regions:
	EC (red line), EH (blue line), EHH (black line).
	{\em Top}: \emt\ curves derived with MCMC method, using \eis\
	and \xrt\ fluxes. {\em Bottom}: \emt\ derived from \xrt\ 
	data through the forward fitting 2T method (see \S\ref{ss:xrt} 
	for details), using photospheric abundances.
	\label{fig:EM_3reg}}
\end{figure}

In Figure~\ref{fig:EM_3reg} we show the emission measure distributions 
obtained for the three regions, using the MCMC algorithm applied to 
\xrt\ and \eis\ data together ({\em top panel}), or forward fitting the 
\xrt\ observations using two temperature components ({\em bottom panel}).
We note that despite the forward fitting 2T model adopts considerable
simplifying assumptions for the underlying emission measure 
distribution the results we obtain with this method are qualitatively 
in good agreement with the MCMC approach which does not impose 
constraints on the shape of the DEM. On the other hand, we note that the
forward fitting 2T model has the advantage of also taking into account the
spatial information, while the MCMC method only uses the information
contained in the integrated fluxes in each passband.
The two hotter regions, EH and EHH, have very similar underlying \emt: 
their cool components have analogous peak temperature 
($\log T [{\rm K}] \sim 6.2-6.3$) and amounts of plasma, whereas the 
hot component of EHH is of weight comparable to the high
temperature component of EH but it appears shifted towards slightly
higher temperatures. For region EC the cool component is characterized
by a peak temperature slightly cooler than the other two regions, 
and in particular its \emt\ falls off faster on the high temperature 
side of the cool peak; the hot component of EC is much weaker than
the other two components, if at all present.

\section{Discussion}
\label{s:discuss}

We have presented the results of a detailed analysis of \hinode\ 
\xrt\ and \eis\ observations of a non-flaring active region to 
diagnose the temperature distribution of the coronal plasma. 
In this work we carried out independent temperature analysis 
from measurements of either instruments, and use the derived
thermal distributions to make prediction for the fluxes
observed with the other instrument and then compare them with 
actual measurements. Then we combine the information from both
instruments and compare the results with the findings based on
\xrt\ or \eis\ data only.
This approach allowed us to explore the limitations of the data 
and analysis methods in providing constraints to the plasma 
temperature distribution, and investigate the cross-calibration
of the two \hinode\ instruments.
With respect to previous works using both \xrt\ and \eis\ data to
study the thermal properties of the coronal plasma 
\citep[e.g.,][]{Landi10}, in this work we provide a 
quantitative cross-check between the diagnostics of the two
instruments.

When using the two datasets separately the derived 
\emt\ curves have overall similar width and peak temperature. 
However, the emission measure derived from \eis\ is systematically
smaller, by a factor $\sim 2-3$, than the emission measure derived from 
\xrt\ fluxes, for each of the three studied regions. We note that 
even if the uncertainties in the derived \emt\ are rather large the 
systematic discrepancy appears to be significant.
We find that the extent of the discrepancy between \xrt\ and \eis\ 
depends in part on the assumptions made for the chemical composition of
the X-ray emitting plasma: it can be mitigated by assuming
abundances intermediate between the typical ``coronal'' abundances
\citep{Feldman92} and the photospheric values \citet{GrevesseSauval},
and it almost disappears when using photospheric abundances.
This interesting finding encourages further investigations to search
for similar evidence in other active regions, in an active region at 
different epochs, and in different kinds of coronal structures (quiet
sun, bright points, coronal holes).
While our study provides no robust constraints on the element
abundances we stress the importance of the assumed values in the
analysis of \hinode\ observations.
The detailed \emt\ reconstruction and the instrument cross-calibration
are significantly affected by the assumed abundances: \eis\ spectra 
allow the determination of ``abundance independent'' \emt\, e.g. 
using exclusively Fe lines \citep{Watanabe07}, but its absolute values 
depend on the abundances; the \xrt\ passbands include significant 
contribution of several elements, both high-FIP and low-FIP, and therefore
the dependence of the \emt\ on abundances is more complicated 
and temperature dependent.

For the analysis of \xrt\ data we use two different methods for 
deriving the emission measure distribution. We find that the 
Markov-chain Monte Carlo method yields results in good qualitative
agreement with the forward fitting 2T method which makes rather 
simplifying assumptions on the emission measure distribution but
has the advantage of making use also of the spatial information.
This finding lends further confidence to the results obtained 
previously by applying this method to the study of \xrt\ observations 
of another active region \citep{Reale09}. 

We find that the \eis\ spectra allow an accurate determination
of the cool ($\log T [$K$] \lesssim 6.5$), most prominent, component,
but when used alone the \eis\ data are unable to constrain the hotter
emission due to  the lack of strong hot lines in the spectra. 
The combination with \xrt\ data provide much tighter constraints
on the emission measure distribution on a wider temperature
range.
Our analysis shows that the studied active region is characterized 
by bulk plasma temperatures of $\sim 2$~MK, which are rather low for 
active region plasma.  The \emt\ of the three selected regions are 
very similar for $\log T [$K$] \lesssim 6.2$, but for the two hotter
regions, EH and EHH, the cool component is broader and with a peak
shifted towards higher T.
The high temperature tail of the emission measure distribution is 
not strongly constrained by the data.  However, the \emt\ derived 
at least for the two hotter regions, EH and EHH, suggests the 
presence of a hotter component ($\log T[{\rm K}] \gtrsim 6.5$)
about two orders of magnitude weaker than the dominant cool
component, in good agreement with the findings of \citet{Reale09}
for another, hotter, active region.

\section{Conclusions}
\label{s:conclusions}

We analyzed \hinode\ \xrt\ multi-filter data and \eis\ spectral 
observations of a non-flaring active region to study the temperature 
distribution of coronal plasma, \emt, and to carry out a detailed 
investigation of the cross-calibration of the two instruments.
We selected three subareas of the active region, and for each of 
them we derived the emission measure distribution \emt\ by:
(1) using \eis\ measured line fluxes; (2) using \xrt\ 
fluxes in six of the instrument's filters; (3) combining the 
datasets for the two instruments. 

We find a good consistency in the qualitative characteristics
-- peak temperature and width of dominant temperature component-- of 
the \emt\ derived with different methods. However, the emission 
measure distributions derived by the two instruments \xrt\ and \eis\
indicate a systematic discrepancy between the two instruments, with
\eis\ data yielding \emt\ consistently smaller, by about a factor 2,
than the \emt\ compatible with \xrt\ data.
We discuss the possible origin of the disagreement and find that
the assumptions for the element abundances significantly influence
the plasma temperature diagnostics. In particular we find that 
a chemical composition intermediate between the usually adopted
coronal abundances by \citet{Feldman92} and the solar photospheric 
abundances improves the comparison between the results obtained
with the two instruments.  When adopting photospheric abundances
  the discrepancy between \eis\ and \xrt\ decreases further. However
  we note that it seems unlikely that the observed plasma, in
  an active region which is not newly emerged, has photospheric
  composition. Furthermore, the used data do not allow a definite
  determination of the abundances, and therefore do not allow us to
  robustly assess of the cross-calibration of the instruments.

One of the main aims of this work was to exploit the complementary 
diagnostics for the X-ray emitting plasma provided by \xrt\ and \eis,
to investigate the presence of hot plasma in non-flaring regions
and test nanoflare heating models.  
We find that the derived \emt\ are characterized by an expected
dominant cool component (typically $\log T[{\rm K}] \sim 6.3$),
and a much weaker amount of plasma at higher temperature. While
the amount of hot plasma is in general in agreement with recent
findings for other non-flaring active regions, and it is compatible
with expectations from nanoflare models, we find that within
the uncertainties these results are not conclusive.

\begin{acknowledgements}
We thank P.\ Grigis, and J.\ Drake, for useful discussions. 
\hinode\ is a Japanese mission developed and launched by ISAS/JAXA, with 
NAOJ as domestic partner and NASA and STFC (UK) as international 
partners. It is operated by these agencies in co-operation with ESA 
and the NSC (Norway).
P.T. and E.E.D. were supported by NASA contract NNM07AB07C to the 
Smithsonian Astrophysical Observatory.
F.R. acknowledges support from Italian Ministero dell'Universit\`a 
e Ricerca and Agenzia Spaziale Italiana (ASI), contract I/015/07/0.
The work of Enrico Landi is supported by the NNH06CD24C, NNH09AL49I
and other NASA grants.
\end{acknowledgements}

\end{document}